\pdfoutput=1
\documentclass[twocolumn,sts,authoryear,preprint]{imsart}

\RequirePackage[OT1]{fontenc}
\usepackage{amsthm,amsfonts,amsmath,amssymb,natbib}
\usepackage{xspace} 
\usepackage{undertilde}
\usepackage{graphicx}
\usepackage{subfigure}
\usepackage{pgf}
\usepackage{pgffor}
\usepgflibrary{plothandlers}
\usepackage{tikz}
\usepackage{verbatim}
\usepackage{algorithm}
\usepackage{bm}

\startlocaldefs
\theoremstyle{definition}
\newtheorem{defn}{Definition}
\newtheorem{examp}{Example}[section]
\endlocaldefs

\begin{document}

\begin{frontmatter}

\title{Inference $\!$for $\!$Graphs $\!$and $\!$Networks: Extending $\!$Classical $\!$Tools $\!$to $\!$Modern $\!$Data}%
\runtitle{Inference for Graphs and Networks}

\begin{aug}
\author{\fnms{Benjamin P.} \snm{Olding}\ead[label=e1]{olding@stat.harvard.edu}}
\and
\author{\fnms{Patrick J.} \snm{Wolfe}\ead[label=e2]{wolfe@stat.harvard.edu}%
\ead[label=u1,url]{http://sisl.seas.harvard.edu}}

\runauthor{B.~P.~Olding and P.~J.~Wolfe}

\address{{\small Benjamin P. Olding is Postdoctoral Research Fellow, Statistics and Information Sciences Laboratory, Harvard University, Oxford Street, Cambridge, Massachusetts 02138, USA\\
\printead{e1}}}

\address{{\small Patrick J. Wolfe is Associate Professor, School of Engineering \& Applied Sciences and Department of Statistics, Harvard University, Oxford Street, Cambridge, Massachusetts 02138, USA\\
\printead{e2}}}

\end{aug}

\begin{abstract}%
Graphs and networks provide a canonical representation of relational data, with massive network data sets becoming increasingly prevalent across a variety of scientific fields.  Although tools from mathematics and computer science have been eagerly adopted by practitioners in the service of network inference, they do not yet comprise a unified and coherent framework for the statistical analysis of large-scale network data.  This paper serves as both an introduction to the topic and a first step toward formal inference procedures.  We develop and illustrate our arguments using the example of hypothesis testing for network structure. We invoke a generalized likelihood ratio framework and use it to highlight the growing number of topics in this area that require strong contributions from statistical science.  We frame our discussion in the context of previous work from across a variety of disciplines, and conclude by outlining fundamental statistical challenges whose solutions will in turn serve to advance the science of network inference.
\end{abstract}



\end{frontmatter}

\section{Introduction}
\label{sec:intro}

Graphs and networks have long been a subject of significant mathematical and scientific interest, deemed worthy of study for their own sake and often associated with scientific data.  However, a
diverse and rapidly growing set of contemporary applications is fast giving rise to massive networks that \emph{themselves} comprise the data set of interest---and to analyze these network data, practitioners in turn require analogs to classical inferential procedures.

While past decades have witnessed a variety of advances in the treatment of graphs and networks as combinatoric or algebraic objects, corresponding advances in formal data analysis have largely failed to keep pace.  Indeed, the development of a successful framework for the statistical analysis of network data requires the repurposing of existing models and algorithms for the specific purpose of inference.  In this paper, we pose the question of how modern statistical science can best rise to this challenge as well as benefit from the many opportunities it presents.  We provide first steps toward formal network inference procedures through the introduction of new tests for network structure, and employ concrete examples throughout that serve to highlight the need for additional research contributions in this burgeoning area.

\subsection{Modern Network Data Sets}
\label{sec:networkData}

Though once primarily the domain of social scientists, a view of networks as data objects is now of interest to researchers in areas spanning biology, finance, engineering, and library science, among others.  \citet{newman03:_structure_and_function_of_comples_networks} provides an extensive review of modern network data sets; other examples of note include mobile phone records, which link customers according to their phone calls~\citep{eagle07:_inferring_social_network_structure_using_mobile_phone_data}; the internet, including both web pages connected by hyperlinks~\citep{adamic00:_power_law_distribution_of_www} and peer-to-peer networks~\citep{willinger06:_unbiased_sampling_peer_to_peer}; electrical power networks, in which local grids are physically connected by long-distance transmission lines~\citep{watts98:_collective_dynamics_of_small_world_networks}; and publication networks, where citations provide explicit links between authors~\citep{desollaprice65:_networks_of_scientific_papers}.

At the same time, other scientific fields are beginning to reinterpret traditional data sets as networks, in order to better understand, summarize, and visualize relationships amongst very large numbers of observations. Examples include protein-protein interaction networks, with isolated pairs of proteins deemed connected if an experiment suggests that they interact~\citep{batada06:_view_of_the_yeast_PPI_network}; online financial transactions, whereupon items are considered to be linked if they are typically purchased together~\citep{jin07:_analysis_of_bidding_networks}; food webs, with species linked by predator-prey relationships~\citep{dunne02:_food_web_structure}; and spatial data sets~\citep{thompson06:_adaptive_web_sampling, ceyhan2007nfr}.

\subsection{Organization and Aims of the Paper}

The above examples attest both to the wide variety of networks encountered in contemporary applications, as well as the multiple expanding literatures on their analysis.  In this paper, we focus on introducing the subject from first principles and framing key inferential questions.  We begin with a discussion of relational data in Section~\ref{sec:relData}, and introduce notation to make the connection to networks precise.  We discuss model specification and inference in Section~\ref{sec:modelSpec}, by way of concrete definitions and examples.  We introduce new ideas for detecting network structure in Section~\ref{sec:DataExample}, and apply them to data analysis by way of formal testing procedures.  In Section~\ref{sec:disc} we discuss open problems and future challenges for large-scale network inference in the key areas of model elicitation, approximate fitting procedures, and issues of data sampling. In a concluding appendix we provide a more thorough introduction to the current literature, highlighting contributions to the field from statistics as well as a variety of other disciplines.

\section{Networks as Relational Data}
\label{sec:relData}

We begin our analysis by making explicit the connection between networks and relational data.   In contrast to data sets that may that arise from pairwise distances or affinities of points in space or time, many modern network data sets are massive, high dimensional, and non-Euclidean in their structure.  We therefore require a way to describe these data other than through purely pictorial or tabular representations---and the notion of cataloging the pairwise relationships that comprise them, with which we begin our analysis, is natural.

\subsection{Relational Data Matrices and Covariates}

Graphs provide a canonical representation of relational data as follows:  Given $n$ entities or objects of interest with pairs indexed by $(i,j)$, we write $i \sim j$ if the $i$th and $j$th entities are related, and $i \nsim j$ otherwise.  These assignments may be expressed by way of an $n \times n$ adjacency matrix $\bm{A}$, whose entries $\{A_{ij}\}$ are nonzero if and only if $i \sim j$. While both the structure of $\bm{A}$ and the field over which its entries are defined depend on the application or specific data set, a natural connection to graph theory emerges in which entities are represented by vertices, and relations by edges; we adopt the informal but more suggestive descriptors ``node'' and ``link,'' respectively.  The \emph{degree} of the $i$th node is in turn defined as $\sum_{j=1}^n A_{ij}$.

In addition, the data matrix $\bm{A}$ is often accompanied by covariates $c(i)$ associated with each node, $i \in \{1,2,\ldots,n\}$.  Example~\ref{ex:simulatedData} below illustrates a case in which these covariates take the form of binary categorical variables.  We shall refer back to these illustrative data throughout Sections~\ref{sec:relData} and~\ref{sec:modelSpec}, and later in Section~\ref{sec:DataExample} will consider a related real-world example: the social network recorded by~\citet{zachary77:_an_information_flow_model_for_conflict_in_small_groups}, in which nodes represent members of a collegiate karate club and links represent friendships, with covariates indicating a subsequent split of the club into two disjoint groups.
\begin{examp}[Network Data Set]\label{ex:simulatedData}
As an example data set, consider the 10-node network defined by data matrix $\bm{A}$ and covariate vector $c$ as follows:
\begin{equation*}
\bm{A} = \begin{pmatrix}
       0 & 0 & 1 & 0 & 1 & 0 & 0 & 0 & 0 & 0 \\
       0 & 0 & 0 & 0 & 0 & 1 & 1 & 0 & 0 & 0 \\
       1 & 0 & 0 & 1 & 1 & 0 & 0 & 0 & 0 & 1 \\
       0 & 0 & 1 & 0 & 0 & 1 & 0 & 1 & 1 & 0 \\
       1 & 0 & 1 & 0 & 0 & 0 & 0 & 0 & 0 & 0 \\
       0 & 1 & 0 & 1 & 0 & 0 & 0 & 0 & 0 & 1 \\
       0 & 1 & 0 & 0 & 0 & 0 & 0 & 0 & 0 & 1 \\
       0 & 0 & 0 & 1 & 0 & 0 & 0 & 0 & 1 & 0 \\
       0 & 0 & 0 & 1 & 0 & 0 & 0 & 1 & 0 & 1 \\
       0 & 0 & 1 & 0 & 0 & 1 & 1 & 0 & 1 & 0
     \end{pmatrix}\!; \,\,\,
c     = \begin{pmatrix}
       1 \\ 0 \\ 0 \\ 0 \\ 1 \\ 0 \\ 1 \\ 1 \\ 1 \\ 0
       \end{pmatrix} \!\text{.}
\end{equation*}
A visualization of the corresponding network is shown in Fig.~\ref{fig:graphExample};
 \begin{figure}[t]
    \center{\includegraphics[width=0.75\columnwidth]{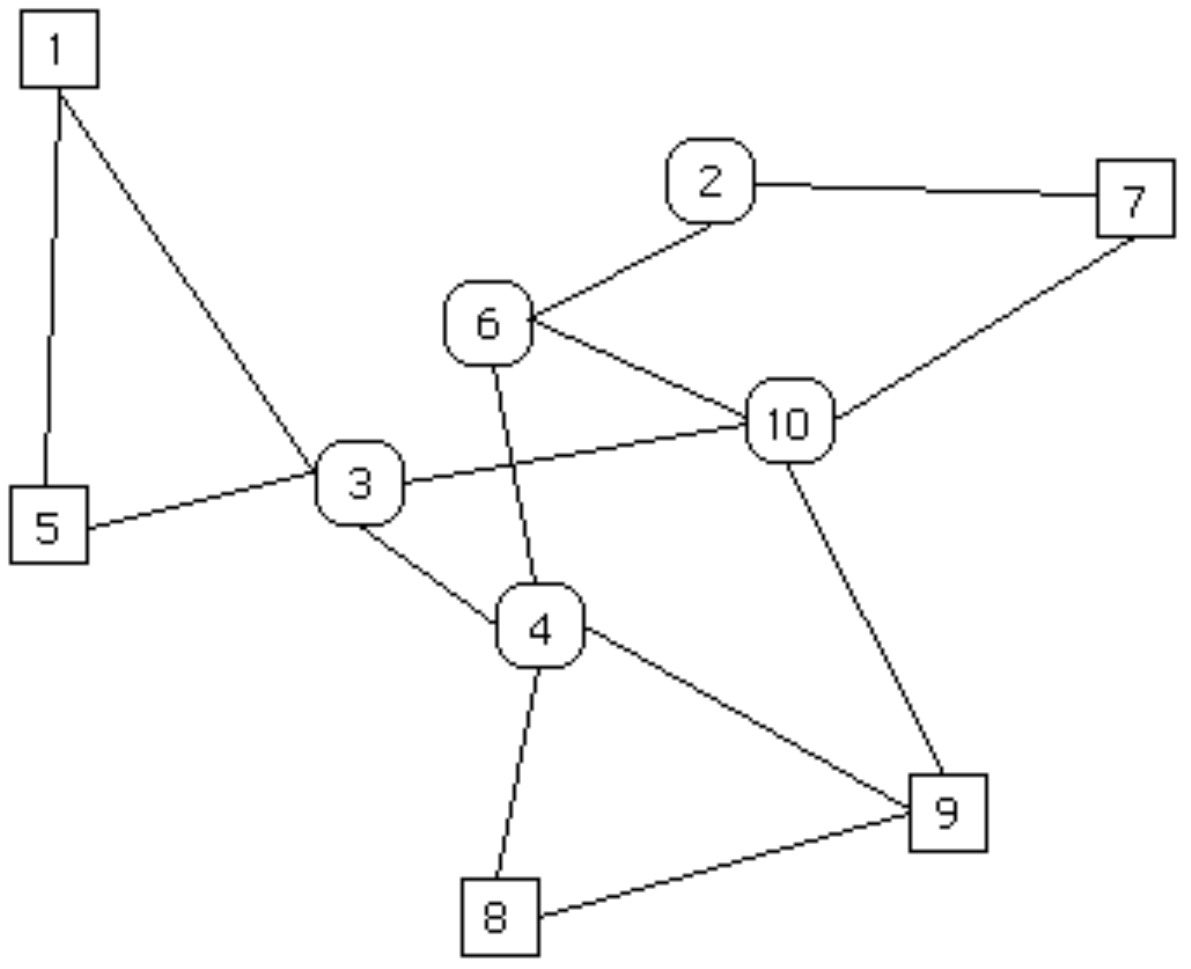}}
    \caption{The network data of Example~\ref{ex:simulatedData}, with nodes indexed by number and binary categorical covariate values by shape.  Note that no Euclidean embedding accompanies the data, making visualization a challenging task for large-scale networks}
    \label{fig:graphExample}
\end{figure}
however, note that as no geometric structure is implied by the data set itself, a pictorial rendering such as this is arbitrary and non-unique.
\end{examp}

In Example~\ref{ex:simulatedData}, categorical covariates $c(i), i \in \{1,2,\ldots,n\}$ are given; however, in almost all network data sets of practical interest, these covariates are latent.  This in turn gives rise to many of the principal questions of network inference---in contrast to the traditional setting of relational data.  Therefore, the issues of network modeling which arise tend to be distinct; as such, classical approaches (e.g., contingency table tests) are directly applicable to network data only in very restricted circumstances.

\subsection{Networks as Distinct From Relational Data}

The main distinction between modern-day network data and classical relational data lies in the requisite computational complexity for inference.  Indeed, the computational requirements of large-scale network data sets are substantial.  With $n$ nodes we associate $\binom{n}{2} = n(n-1)/2$ symmetric relations; beyond this quadratic scaling, latent covariates give rise to a variety of combinatorial expressions in $n$.  Viewed in this light, methods to determine relationships amongst \emph{subsets} of nodes can serve as an important tool to ``coarsen'' network data.  In addition to providing a lower-dimensional summary of the data, such methods can serve to increase the computational efficiency of subsequent inference procedures by enabling data reduction and smoothing.  The general approach is thus similar to modern techniques for high-dimensional Euclidean data, and indeed may be viewed as a clustering of nodes into groups.

From a statistical viewpoint, this notion of subset relations can be conveniently described by a $k$-ary categorical covariate, with $k$ specifying the (potentially latent) model order.  By incorporating such a covariate into the probability model for the data adjacency matrix $\bm{A}$, the ``structure'' of the network can be directly tested if this covariate is observed, or instead inferred if latent.  It is easily seen that the cardinality of the resultant model space is exponential in the number of nodes $n$; even if the category sizes themselves are given, we may still face a combinatorial inference problem.  Thus, even a straightforwardly-posed hypothesis test for a relatively simple model can easily lead to cases where exact inference procedures are intractable.

\section{Model Specification and Inference}\label{sec:modelSpec}

Fields such as probability, graph theory, and computer science have each posited specific models which can be applied to network data; however, when appealing to the existing literature, it is often the case that neither the models nor the analysis tools put forward in these contexts have been developed specifically for inference.  In this section, we introduce two basic network models and relate them to classical statistics.  The first such model consists of nodes that are probabilistically exchangeable, whereas the second implies group structure via latent categorical covariates.  Inferring relationships amongst groups of nodes from data in turn requires the standard tools of statistics, including parameter estimation and hypothesis testing.  We provide examples of such procedures below,
illustrating their computational complexity, and introduce corresponding notions of approximate inference.

\subsection{Erd\"{o}s-R\'{e}nyi: A First Illustrative Example}
\label{sec:E-R}

We begin by considering one of the simplest possible models from random graph theory, attributed to~\citet{erdos59:_on_random_graphs} and~\citet{gilbert59:_random_graphs}, and consisting of pairwise links that are generated independently with probability $p$.  Under this model, all nodes are exchangeable; it is hence appropriate to describe instances in which no group structure (by way of categorical covariates) is present.  In turn, we shall contrast this with an explicit model for structure below.

Adapted to the task of modeling undirected network data, the Erd\"{o}s-R\'{e}nyi model may be expressed as a sequence of $\binom{n}{2}$ Bernoulli trials corresponding to off-diagonal elements of the adjacency matrix $\bm{A}$.

\begin{defn}[Erd\"{o}s-R\'{e}nyi Model]\label{def:ER}
Let $n > 1$ be integral and fix some $p \in [0,1]$.  The Erd\"{o}s-R\'{e}nyi random graph model corresponds to matrices $\bm{A} \in \{0,1\}^{n\times n}$ defined element-wise as
\begin{align*}
\forall\, i,j \, \in \{1,2,\ldots,n\}: i < j,  \quad &
A_{ij} \overset{\text{iid}}{\sim} \operatorname{Bernoulli}(p); \\
& A_{ji}=A_{ij}, \,\, A_{ii} = 0  \text{.}
\end{align*}
\end{defn}
Erd\"{o}s-R\'{e}nyi thus provides a one-parameter model yielding independent and identically distributed binary random variables representing the absence or presence of pairwise links between nodes; as this binary relation is symmetric, we take $A_{ji}=A_{ij}$.  The additional stipulation $A_{ii}=0$ for all $i$ implies that our relation is also irreflexive; in the language of graph theory, the corresponding (undirected, unweighted) graph is said to be \emph{simple}, as it exhibits neither multiple edges nor self-loops.  The event $i \sim j$ is thus a $\operatorname{Bernoulli}(p)$ random variable for all $i \neq j$, and it follows that the degree $\sum_{j=1}^n A_{ij}$ of each network node is a $\operatorname{Binomial}(n-1,p)$ random variable.

Fitting the parameter $p$ is straightforward; the maximum likelihood estimator (MLE) corresponds to the sample proportion of observed links:
\begin{equation*}
\widehat{p} := \frac{1}{\textstyle \binom{n}{2}} \sum_{i<j} A_{ij} = \frac{1}{n(n-1)} \sum_{i=1}^n \sum_{j=1}^n A_{ij} \text{.}
\end{equation*}
Example~\ref{ex:simulatedData}, for instance, yields $\widehat{p} = 14/45$.

Given a relational data set of interest, we can test the agreement of data in $\bm{A}$ with this model by employing an appropriately selected test statistic.  If we wish to test this uniformly generic model with respect to the notion of network structure, we may explicitly define an alternate model and appeal to the classical Neyman-Pearson testing framework.

In this vein, the Erd\"{o}s-R\'{e}nyi model can be generalized in a natural way to capture the notion of local rather than global exchangeability: we simply allow Bernoulli parameters to depend on
$k$-ary categorical covariates $c(i)$ associated with each node $i \in \{1,2,\ldots,n\}$, where the $k \leq n$ categories represent groupings of nodes.  Formally we define
\begin{equation*}
c \in \mathbb{Z}_k^n; \quad c(i) : \{1,2,\ldots,n\} \mapsto \mathbb{Z}_k \text{,}
\end{equation*}
and a set of $\binom{k+1}{2}$ distinct Bernoulli parameters governing link probabilities within and between these categories, arranged into a $k \times k$ symmetric matrix and indexed as $p_{c(i) c(j)}$ for $i,j \in \{1,2,\ldots,n\}$.

In the case of binary categorical covariates, we immediately obtain a formulation of~\citet{holland81:_exponential_family_for_directed_graphs}, the simplest example of a so-called \emph{stochastic block model}.  In this network model, pairwise links between nodes correspond again to Bernoulli trials, but with a parameter chosen from the set $\{p_{00}, p_{01}, p_{11}\}$ according to binary categorical covariates associated with the nodes in question.

\begin{defn}[Simple Stochastic Block Model]\label{def:dyadInd}
Let $c \in \{0,1\}^n$ be a binary $n$-vector for some integer $n>1$, and fix parameters $p_{00}, p_{01}, p_{11} \in [0,1]$.  Set $p_{10}=p_{01}$; the model then corresponds to matrices $\bm{A} \in \{0,1\}^{n\times n}$ defined element-wise as
\begin{align*}
\forall\, i,j \in \{1,2,\ldots,n\}: i < j,  \, & A_{ij} \sim \operatorname{Bernoulli}(p_{c(i)c(j)}); \\
& A_{ji}=A_{ij}, \,\, A_{ii} = 0  \text{.}
\end{align*}
\end{defn}

\vspace{-0.5\baselineskip}%
If the vector of covariates $c$ is given, then finding the maximum-likelihood parameter estimates $\{\widehat{p}_{00}, \widehat{p}_{01}, \widehat{p}_{11}\}$ is trivial after a re-ordering of nodes via permutation similarity: For any $n \times n$ permutation matrix $\bm{\Pi}$, the adjacency matrices $\bm{A}$ and $\bm{\Pi}\bm{A}\bm{\Pi}'$ represent isomorphic graphs, the latter featuring permuted rows and columns of the former.  If $\bm{\Pi}$ re-indexes nodes according to their categorical groupings, then we may define a conformal partition
\begin{equation*}
\bm{\Pi}\bm{A}\bm{\Pi}' = \begin{pmatrix}\bm{A_{00}} & \bm{A_{01}} \\ \bm{A_{01}}' & \bm{A_{11}} \end{pmatrix}
\end{equation*}
that respects this ordering, such that exchangeability is preserved within---but not across---submatrices $\bm{A_{00}}$ and $\bm{A_{11}}$.  We may then simply compute sample proportions corresponding to each submatrix $\{\bm{A_{00}}, \bm{A_{01}}, \bm{A_{11}}\}$ to yield $\{\widehat{p}_{00}, \widehat{p}_{01}, \widehat{p}_{11}\}$.

Note that by construction, submatrices $\bm{A_{00}}$ and $\bm{A_{11}}$ yield subgraphs that are themselves Erd\"{o}s-R\'{e}nyi, and are said to be \emph{induced} by the two respective groups of categorical covariates.  Nonzero entries of $\bm{A_{01}}$ are said to comprise the \emph{edge boundary} between these two induced subgraphs; indeed, the matrix obtained by setting all entries of $\bm{A_{00}}$ and $\bm{A_{11}}$ to zero yields in turn a \emph{bipartite} graph whose vertices can be partitioned according to their binary covariate values.

The following example illustrates these concepts using the simulated data of Example~\ref{ex:simulatedData}.

\begin{examp}[Similarity and Subgraphs]\label{ex:inducedSubgraph}
Let the 10-node network of Example~\ref{ex:simulatedData} be subject to an isomorphism that re-orders nodes according to the two groups defined by their binary covariate values, and define the permuted covariate vector $\tilde{c}$ and permutation-similar data matrix $\bm{\widetilde{A}}$ as follows:
\begin{align*}
\tilde{c}' & = \!\begin{pmatrix}
       0 & 0 & 0 & 0 & 0 & 1 & 1 & 1 & 1 & 1
       \end{pmatrix}; \\
\bm{\widetilde{A}}
& \!=\!
\begin{pmatrix}
       0 & 0 & 0 & 1 & 0\,\vline &\!\! 0 & 0 & 1 & 0 & 0 \\
       0 & 0 & 1 & 0 & 1\,\vline &\!\! 1 & 1 & 0 & 0 & 0 \\
       0 & 1 & 0 & 1 & 0\,\vline &\!\! 0 & 0 & 0 & 1 & 1 \\
       1 & 0 & 1 & 0 & 1\,\vline &\!\! 0 & 0 & 0 & 0 & 0 \\
       0 & 1 & 0 & 1 & 0\,\vline &\!\! 0 & 0 & 1 & 0 & 1 \\
       \hline
       0 & 1 & 0 & 0 & 0\,\vline &\!\! 0 & 1 & 0 & 0 & 0 \\
       0 & 1 & 0 & 0 & 0\,\vline &\!\! 1 & 0 & 0 & 0 & 0 \\
       1 & 0 & 0 & 0 & 1\,\vline &\!\! 0 & 0 & 0 & 0 & 0 \\
       0 & 0 & 1 & 0 & 0\,\vline &\!\! 0 & 0 & 0 & 0 & 1 \\
       0 & 0 & 1 & 0 & 1\,\vline &\!\! 0 & 0 & 0 & 1 & 0
     \end{pmatrix}
      \!\!=\!\!
\begin{pmatrix}\bm{\widetilde{A}_{00}} & \!\!\!\bm{\widetilde{A}_{01}} \\ {\bm{\widetilde{A}_{01}}}' & \!\!\!\bm{\widetilde{A}_{11}} \end{pmatrix}\!\!\text{.}
\end{align*}
Figure~\ref{fig:graphExampleSubgraph}
\begin{figure}[t]
    \center{\includegraphics[width=0.75\columnwidth]{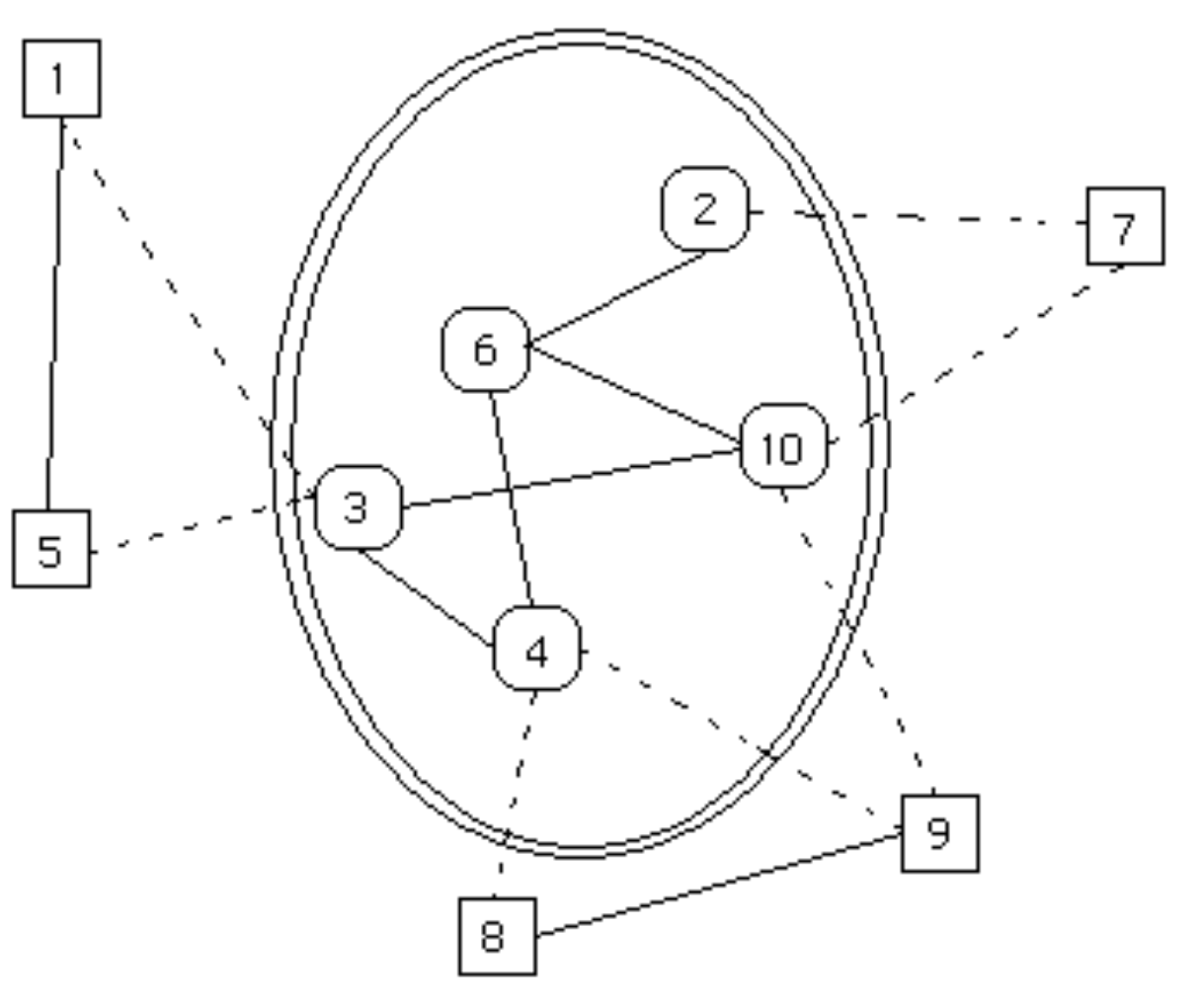}}
    \caption{Subgraphs based on the binary covariates of Example~\ref{ex:simulatedData}, again represented graphically by node shape.  The conformal partition of Example~\ref{ex:inducedSubgraph} implies two induced subgraphs: solid lines inside the ellipse are links represented in submatrix $\bm{\widetilde{A}_{00}}$, while those outside it appear as links in $\bm{\widetilde{A}_{11}}$.
    The remaining links, shown as dashed lines, correspond to values of 1 in submatrix $\bm{\widetilde{A}_{01}}$ and comprise the associated edge boundary}
    \label{fig:graphExampleSubgraph}
\end{figure}
illustrates the corresponding subgraphs using the visualization of Fig.~\ref{fig:graphExample}; assuming a simple stochastic block model in turn leads to the following maximum-likelihood parameter estimates:
\begin{equation*}
\widehat{p}_{00} = \frac{5}{10}; \quad \widehat{p}_{01} = \frac{7}{25}; \quad \widehat{p}_{11} = \frac{2}{10}\text{.}
\end{equation*}
\end{examp}

Example~\ref{ex:inducedSubgraph} illustrates the ease of model fitting when binary-valued covariates are known; the notion of permutation similarity plays a similar role in the case of $k$-ary covariates.

\subsection{Approximate Inference}
\label{sec:ApproximateInference}

The careful reader will have noted that in the case of known categorical covariates, examples such as those above
can be expressed as contingency tables---a notion we revisit in Section~\ref{sec:DataExample} below---and hence may admit exact inference procedures.  However, if covariates are latent, then an appeal to maximum-likelihood estimation induces a combinatorial optimization problem; in general, no fast algorithm is known for likelihood maximization over the set of covariates and Bernoulli parameters under the general $k$-group stochastic block model.

The principal difficulty arises in maximizing the $n$-dimensional $k$-ary covariate vector $c$ over an exponentially large model space; estimating the $\binom{k+1}{2}$ associated Bernoulli parameters then proceeds in exact analogy to Example~\ref{ex:inducedSubgraph} above.  The following example illustrates the complexity of this inference task.

\begin{examp}[Permutation and Maximization]\label{ex:likelihoodMax}
Consider a 100-node network generated according to the stochastic block model of Definition~\ref{def:dyadInd}, with each group of size 50 and $p_{00} = p_{11} = 1/2$, $p_{01} = 0$.  Figure~\ref{fig:matrix_visualization}
 \begin{figure}[t]
    \center{\subfigure[Simulated data matrix $\bm{A}$]
      {\label{subfig:matrix_visualization:unordered}\includegraphics[width=0.65\columnwidth]{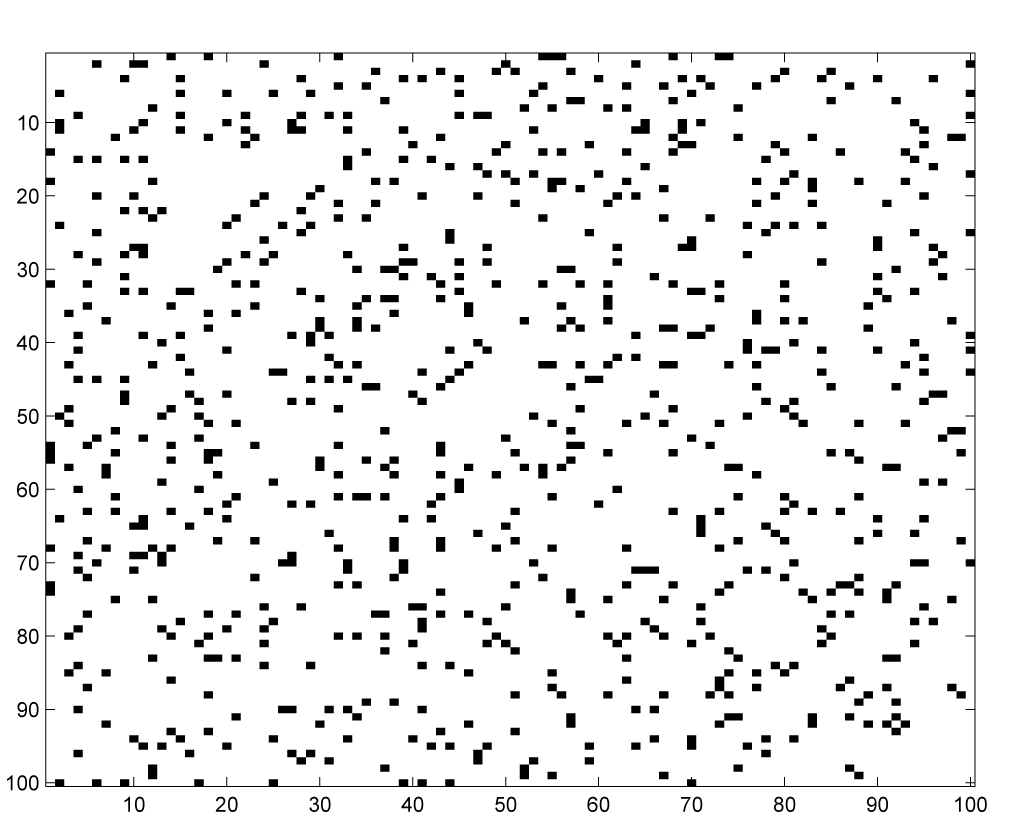}}
      \subfigure[Transformed data $\bm{\Pi}\bm{A}\bm{\Pi}'$]
      {\label{subfig:matrix_visualization:ordered}\includegraphics[width=0.65\columnwidth]{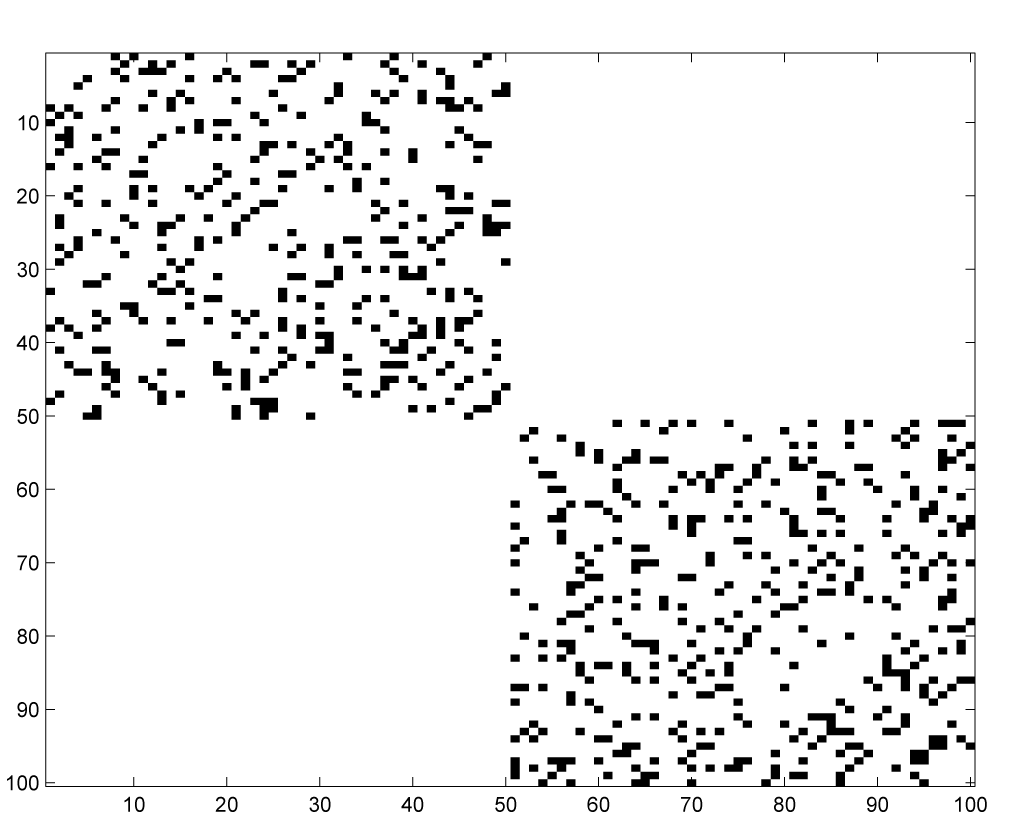}}}
    \caption{Representations $\bm{A}$ and $\bm{\Pi}\bm{A}\bm{\Pi}'$ of data drawn from the stochastic block model of Example~\ref{ex:likelihoodMax}, corresponding to isomorphic graphs (black boxes denote links).  Though $p_{01} = 0$, only a small subset of permutation similarity transformations $\bm{\Pi}(\cdot)\bm{\Pi}'$ will reveal the disconnected nature of this network}
    \label{fig:matrix_visualization}
\end{figure}
shows two permutation-similar adjacency matrices, $\bm{A}$ and $\bm{\Pi}\bm{A}\bm{\Pi}'$, that correspond to isomorphic graphs representing this network; inferring the vector $c$ of binary categorical covariates from data $\bm{A}$ in Fig.~\ref{subfig:matrix_visualization:unordered} is equivalent to finding a permutation similarity transformation $\bm{\Pi}\bm{A}\bm{\Pi}'$ that reveals the distinct division apparent in Fig.~\ref{subfig:matrix_visualization:ordered}.
\end{examp}

Given the combinatorial nature of this problem, it is clear that fitting models to real-world network data can quickly necessitate approximate inference.  To this end, Example~\ref{ex:likelihoodMax} motivates an important means of exploiting algebraic properties of network adjacency structure: the notion of a \emph{graph spectrum}.  Eigenvalues associated with graphs reveal several of their key properties~\citep{chung1997sgt} at a computational cost that scales as the cube of the number of nodes, offering an appealing alternative in cases where exact solutions are of exponential complexity.

As the adjacency matrix $\bm{A}$ itself fails to be positive semidefinite, the spectrum of a labeled graph is typically defined via a Laplacian matrix $\bm{L}$ as follows.

\begin{defn}[Graph Laplacian]
Let $i \sim j$ denote a symmetric adjacency relation defined on an $n$-node network.  An associated $n \times n$ symmetric, positive-semidefinite matrix $\bm{L}$ is called a graph Laplacian if, for all $i,j \in \!\{1,2\ldots,n\}:i \neq j$, we have
\begin{equation*}
\bm{L}: \quad
\begin{cases}
L_{ij} < 0 & \text{if $i \sim j$,} \\
L_{ij} = 0 & \text{if $i \nsim j$.}
\end{cases}; \quad L_{ji} = L_{ij} \text{.}
\end{equation*}
\end{defn}

Note that the diagonal of $\bm{L}$ is defined only implicitly, via the requirement of positive-semidefiniteness; a typical diagonally-dominant completion termed the \emph{combinatorial Laplacian} takes $\bm{L} = \bm{D} - \bm{A}$, where $\bm{D}$ is a diagonal matrix of node degrees such that $D_{ii} = \sum_{j=1}^n A_{ij}$.
An important result is that the dimension of the kernel of $\bm{L}$ is equal to the number of connected components of the corresponding graph; hence $p_{01} = 0$ implies in turn that at least two eigenvalues of $\bm{L}$ will be zero in Example~\ref{ex:likelihoodMax} above.

Correspondingly, \citet{fiedler73:_algebraic_connectivity_of_graphs} termed the second-smallest eigenvalue of the combinatorial Laplacian the \emph{algebraic connectivity} of a graph, and recognized that positive and negative entries of the corresponding eigenvector (the ``Fiedler vector'') define a partition of nodes that nearly minimizes the number of edge removals needed to disconnect a network.  In fact, in the extreme case of two equally sized, disconnected subgraphs---as given by Example~\ref{ex:likelihoodMax}---this procedure exactly maximizes the likelihood of the data under a two-group stochastic block model; more generally, it provides a means of approximate inference that we shall return to in Section~\ref{sec:DataExample} below.

As reviewed by \citet{luxburg2007tsc}, the observation of Fiedler was later formalized as an algorithm termed spectral bisection~\citep{pothen90:_partitioning_sparse_matrices}, and indeed leads to the more general notion of \emph{spectral clustering}~\citep{vonluxburg2008csc}.  This remains an active area of research in combinatorics and theoretical computer science, where a simple stochastic block model with $p_{00},p_{11} > p_{01}$ is termed a ``planted partition'' model~\citep{bollobas2004mcr}.

\section{Testing for Network Structure}
\label{sec:DataExample}

Identifying some degree of structure within a network data set is an important prerequisite to formal statistical analysis.  Indeed, if all nodes of a network are truly unique and do not admit any notion of grouping, then the corresponding data set---no matter how large---is really only a single observation.  On the other hand, if every node can be considered independent and exchangeable under an assumed model, then depicting the data set as a network is unhelpful: the data are best summarized as $n$ independent observations of nodes whose connectivity structure is uninformative.

In this section we invoke a formal hypothesis testing framework to explore the notion of detecting network structure in greater detail, and propose new approaches that are natural from a statistical point of view but have thus far failed to appear in the literature.  To illustrate these ideas we apply three categories of tests to a single data set---that of Section~\ref{ex:Zachary} below---and in turn highlight a number of important topics for further development.

\subsection{The Zachary Karate Data}\label{ex:Zachary}
\begin{figure}
    \center{\includegraphics[width=0.8\columnwidth]{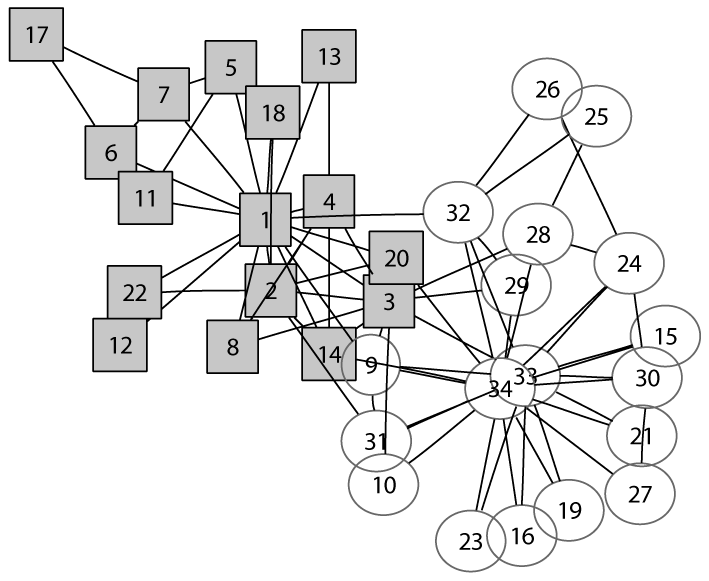}}
    \caption{Visualization of the Zachary karate data of Section~\ref{ex:Zachary}.   Nodes are numbered and binary categorical covariate values, reflecting the subsequent group split, are indicated by shape}
    \label{fig:karate_club}
\end{figure}

\citet{zachary77:_an_information_flow_model_for_conflict_in_small_groups} recorded friendships between 34 members of a collegiate karate club that subsequently split into two groups of size 16 and 18.  These data are shown in Fig.~\ref{fig:karate_club},
with inter- and intra-group links given in Table~\ref{table:chi2_table}.
\begin{table}
  \center{%
  \caption{Contingency table specifying counts of intra- and inter-subgroup links for the data of Section~\ref{ex:Zachary}}
  \label{table:chi2_table}
\begin{tabular}{cccc}
\multicolumn{1}{c}{Counts} & \multicolumn{1}{c}{\# Links} & \multicolumn{1}{c}{\# No Links} & \multicolumn{1}{c}{Total}\\
Intra-subgroup: $0$--$0$ & $33$ & $87$ & $120$\\
Inter-subgroup: $0$--$1$ & $10$ & $278$ & $288$\\
Intra-subgroup: $1$--$1$ & $35$ & $118$ & $153$\\
Total & $78$ & $483$ & $561$\\
\end{tabular}
  }
\end{table}
The network consists of 78 links, with degree sequence (ordered in accordance with the node numbering of Fig.~\ref{fig:karate_club}) given by
\begin{align*}
( & 16, 9, 10, 6, 3, 4, 4, 4, 5, 2, 3, 1, 2, 5, 2, 2, 2,
\\ & 2, 2, 3, 2, 2, 2, 5, 3, 3, 2, 4, 3, 4, 3, 6, 13, 17) \text{,}
\end{align*}
and corresponding sample proportion of observed links given by $\hat{p} = 78/\binom{34}{2} = 78/561$.

Sociologists have interpreted the data of Zachary not only as evidence of network structure in this karate club, but also as providing binary categorical covariate values through an indication of the subsequent split into two groups, as per Fig.~\ref{fig:karate_club}. This in turn provides us with an opportunity to test various models of network structure---including those introduced in Section~\ref{sec:modelSpec}---with respect to ground truth.

\subsection{Tests with Known Categorial Covariates}
\label{sec:NetworkStrutureDetection}

We begin by posing the question of whether or not the most basic Erd\"{o}s-R\'{e}nyi network model of Definition~\ref{def:ER}---with each node being equally likely to connect to any other node---serves as a good description of the data, given the categorical variable of observed group membership.  The classical evaluation of this hypothesis comes via a contingency table test. 

\begin{examp}[Contingency Table Test]\label{ex:ContTable}
Consider the data of Section~\ref{ex:Zachary}.  When categorical covariates are known, a contingency table test for independence between rows and columns may be performed according to the data shown in Table~\ref{table:chi2_table}.  The Pearson test statistic $T_{\chi^2}$ in this case evaluates to over $47$, and with only $2$ degrees of freedom, the corresponding p-value for these data is less than $10^{-3}$.
\end{examp}

In this case, the null hypothesis---that the Erd\"{o}s-R\'{e}nyi model's sole Bernoulli parameter can be used to describe both inter- and intra-subgroup connection probabilities---can clearly be rejected.

As in the case of \citet{zheng06:_how_many_people_do_you_know_in_prison} and others, this $\chi^2$ approach has been generally used to reject an Erd\"{o}s-R\'{e}nyi null when given network data include a categorical covariate for each node.  (A cautionary reminder is in order: employing this method when covariates are inferred from data corresponds to a misuse of maximally selected statistics~\citep{altman94:_dangers_using_optimal_cutpoints}.)  Of course, in cases where it is computationally feasible, we may instead use simulation to determine the exact distribution of any chosen test statistic $T$ under whichever null model is assumed.

\subsection{The Case of Latent Categorial Covariates}

The Erd\"{o}s-R\'{e}nyi model of Definition~\ref{def:ER} clearly implies a lack of network structure through its nodal exchangeability properties, thus supporting its use as a null model in cases such as Example~\ref{ex:ContTable} and those described above.  In contrast, the partial exchangeability exhibited by the stochastic block model of Definition~\ref{def:dyadInd} suggests its use as an alternate model that explicitly exhibits network structure.  To this end, the usual Neyman-Pearson logic implies the adoption of a generalized likelihood ratio test statistic:
\begin{align*}
T_{\mathit{LR}} &= \frac{\sup\limits_{p}\prod\limits_{i>j} \mathbb{P}(A_{ij} \,;p)} {\max\limits_{c}\sup\limits_{p_{00},p_{01},p_{11}}\prod\limits_{i>j} \mathbb{P}(A_{ij} \, ; p_{00},p_{01},p_{11},c(i),c(j))}\\
&=\frac{\prod\limits_{i>j}\widehat{p}^{\,A_{ij}}\left(1-\widehat{p}\,\right)^{1-A_{ij}}}{\max\limits_{c}\sup\limits_{p_{00},p_{01},p_{11}}\prod\limits_{i>j}(p_{c(i)c(j)})^{A_{ij}}(1-p_{c(i)c(j)})^{1-A_{ij}}}\text{.}
\end{align*}

As we have seen in Section~\ref{sec:ApproximateInference}, however, maximizing the likelihood of the covariate vector $c\in\{0,1\}^n$ in general requires an exhaustive search.  Faced with the necessity of approximate inference, we recall that the spectral partitioning algorithms outlined earlier in Section~\ref{sec:ApproximateInference} provide an alternative to exact likelihood maximization in $c$.  The resultant test statistic $T_{\widehat{\mathit{LR}}}$ is computationally feasible, though with reduced power, and to this end way may test the data of Section~\ref{ex:Zachary} as follows.

\begin{examp}[$\!$Generalized $\!$Likelihood $\!$Ratio $\!$Test$\!$]\label{ex:ER-GLRT}
Let $T_{\mathit{LR}}$ be the test statistic associated with a generalized likelihood ratio test of Erd\"{o}s-R\'{e}nyi versus a two-group stochastic block model, and $T_{\widehat{\mathit{LR}}}$ correspond to an approximation obtained by spectral partitioning in place of the maximization over group membership.  For the data of Section~\ref{ex:Zachary}, simulation yields a corresponding p-value of less than $10^{-3}$ with respect to $T_{\widehat{\mathit{LR}}}$, with Fig.~\ref{fig:ER_ROC}
\begin{figure}
    \center{\input{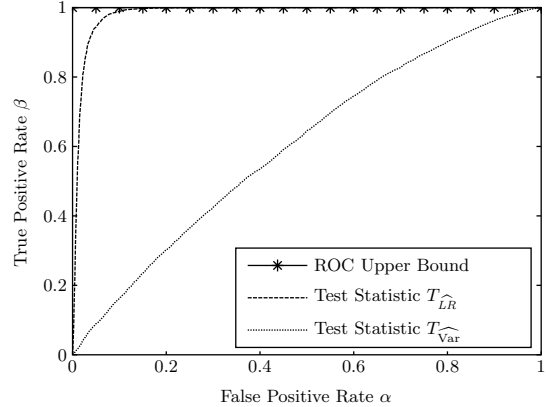}}
    \caption{Receiver operating characteristic (ROC) curves corresponding to tests of the data of Section~\ref{ex:Zachary}, with Erd\"{o}s-R\'{e}nyi null and two-group stochastic block model alternate.  Test statistics $T_{\widehat{\mathit{LR}}}$ and $T_{\widehat{\operatorname{Var}}}$ were calculated via simulation, with the ROC upper bound obtained using knowledge of the true group membership for each node}
    \label{fig:ER_ROC}
\end{figure}
confirming the power of this test.
\end{examp}

Our case study has so far yielded reassuring results.  However, a closer look reveals that selecting appropriate network models and test statistics may require more careful consideration.
\begin{examp}[Degree Variance Test]\label{ex:DegVarTest}
Suppose we adopt instead the test statistic of \citet{snijders81:_the_degree_variance}:
\begin{equation*}
T_{\widehat{\operatorname{Var}}} =
\frac{1}{n-1}\sum_{i=1}^n\Big(\sum_{j=1}^n A_{ij} - \frac{1}{n}\sum_{i=1}^n\sum_{j=1}^nA_{ij}\Big)^2
\text{,}
\end{equation*}
the sample variance of the observed degree sequence $\sum_{j=1}^n A_{ij}$.  A glance at the data of Section~\ref{ex:Zachary} indicates the poor fit of an Erd\"{o}s-R\'{e}nyi null, and indeed simulation yields a p-value of less than $10^{-3}$.  Figure~\ref{fig:ER_ROC}, however, reveals that $T_{\widehat{\operatorname{Var}}}$ possesses very little power.
\end{examp}

This dichotomy between a low p-value, and yet low test power, highlights a limitation of the models exhibited thus far: in each case, both the expected degree sequence and the corresponding node connectivity properties are determined by exactly the same set of model parameters.  In this regard, test statistics depending on the data set only through its degree sequence can prove quite limiting, as the difference between the two models under consideration lies entirely in their node connectivity properties, rather than the heterogeneity of their degree sequences.

Indeed, significant degree variation is a hallmark of many observed network data sets, the data of Section~\ref{ex:Zachary} included; sometimes certain nodes are simply more connected than others.  In order to conclude that rejection of a null model necessarily implies the presence of network structure expressed through categorical covariates, a means of allowing for heterogenous degree sequences must be incorporated into the null as well as the alternate.

\subsection{Decoupling Degree Sequence \& Connectivity}
\label{sec:_two_fixed_degree_models}

An obvious way to decouple properties of the degree sequence from those of connectivity is to restrict the model space to \emph{only} those networks exhibiting the observed degree sequence.  However, simulation of such graphs becomes nontrivial when they are restricted to be simple (i.e., without multiple edges or self-loops), thus rendering the test calculations of Section~\ref{sec:NetworkStrutureDetection} more difficult to achieve in practice.  Correspondingly, such fixed-degree models have remained largely absent from the literature to date.

Recent advances in graph simulation methods, however, help to overcome this barrier~\citep{viger2005eas, blitzstein06:_a_sequential_importance_sampling_algorithm_for_random_graphs}.  The importance sampling approach of \citet{blitzstein06:_a_sequential_importance_sampling_algorithm_for_random_graphs} enables us here to test the data set of Section~\ref{ex:Zachary} using fixed-degree models that match its observed degree sequence.  Although the corresponding normalizing constants cannot be computed in closed form, we may specify a proposal distribution, draw samples, and calculate unnormalized importance weights.

\begin{examp}[Fixed-Degree Test]\label{ex:FDTest}
Consider the set of all simple graphs featuring an observed degree sequence, and define a null model under which each of these graphs is equally likely.  As an alternate model, let each graph be weighted in proportion to its likelihood under the two-group stochastic block model of Definition~\ref{def:dyadInd}; in this case the normalizing constant will depend on parameters $p_{00}$, $p_{01}$, and $p_{11}$.  The corresponding fixed-degree generalized likelihood ratio test statistic $T_{\mathit{LR-FD}}$ is given in analogy to Example~\ref{ex:ER-GLRT} by
\begin{equation*}
\frac{1}{\max\limits_{c}\sup\limits_{p_{00},p_{01},p_{11}}\prod\limits_{i>j} \mathbb{P}(A_{ij} \, ; p_{00},p_{01},p_{11},c(i),c(j))} \text{.}
\end{equation*}

Just as before, calculation of $T_{\mathit{LR-FD}}$ requires a combinatorial search over group assignments $c$; moreover, the fixed-degree constraint precludes an analytical $\sup$ operation over parameters $p_{00}$, $p_{01}$, and $p_{11}$. We therefore define an approximation $T_{\widehat{\mathit{LR}}\mathit{-FD}}$ employing spectral partitioning in place of the maximization over group membership, and substituting the analytical $\sup$ under two-group stochastic block likelihood for the exact $\sup$ operation.  The substantial power of this test for the data of Section~\ref{ex:Zachary} is visible in Fig.~\ref{fig:FD_ROC};
\begin{figure}
    \center{
\begin{tikzpicture} [scale=0.75]
  \begin{pgfscope}
    \definecolor{matfig2pgf_color}{rgb}{1,1,1}\pgfsetfillcolor{matfig2pgf_color}
    \pgfpathrectangle{\pgfpoint{1.41111cm}{1.05833cm}}{\pgfpoint{8.30667cm}{6.15944cm}}
    \pgfusepath{fill}
  \end{pgfscope}
  \begin{pgfscope}
    \pgfsetlinewidth{0.5pt}
    \foreach \x in {1.41111,3.07244,4.73378,6.39511,8.05644,9.71778}
    {
      \pgfpathmoveto{\pgfpoint{\x cm}{1.05833cm}}\pgfpathlineto{\pgfpoint{\x cm}{1.11993cm}}
      \pgfpathmoveto{\pgfpoint{\x cm}{7.21778cm}}\pgfpathlineto{\pgfpoint{\x cm}{7.15618cm}}
    }
    \foreach \y in {1.05833,2.29022,3.52211,4.754,5.98589,7.21778}
    {
      \pgfpathmoveto{\pgfpoint{1.41111cm}{\y cm}}\pgfpathlineto{\pgfpoint{1.49418cm}{\y cm}}
      \pgfpathmoveto{\pgfpoint{9.71778cm}{\y cm}}\pgfpathlineto{\pgfpoint{9.63471cm}{\y cm}}
    }
    \pgfusepath{stroke}
  \end{pgfscope}
  \begin{pgfscope}
    \pgfsetlinewidth{0.5pt}
    \pgfpathrectangle{\pgfpoint{1.41111cm}{1.05833cm}}{\pgfpoint{8.30667cm}{6.15944cm}}
    \pgfusepath{stroke}
  \end{pgfscope}
    \pgftext[x=1.41111cm,y=0.958333cm,top]{$0$}
    \pgftext[x=3.07244cm,y=0.958333cm,top]{$0.01$}
    \pgftext[x=4.73378cm,y=0.958333cm,top]{$0.02$}
    \pgftext[x=6.39511cm,y=0.958333cm,top]{$0.03$}
    \pgftext[x=8.05644cm,y=0.958333cm,top]{$0.04$}
    \pgftext[x=9.71778cm,y=0.958333cm,top]{$0.05$}
    \pgftext[x=1.31111cm,y=1.05833cm,right]{$0.9$}
    \pgftext[x=1.31111cm,y=2.29022cm,right]{$0.92$}
    \pgftext[x=1.31111cm,y=3.52211cm,right]{$0.94$}
    \pgftext[x=1.31111cm,y=4.754cm,right]{$0.96$}
    \pgftext[x=1.31111cm,y=5.98589cm,right]{$0.98$}
    \pgftext[x=1.31111cm,y=7.21778cm,right]{$1$}
  \makeatletter\ifpgf@draftmode\makeatother\else
  \begin{pgfscope}
    \pgfpathrectangle{\pgfpoint{1.41111cm}{1.05833cm}}{\pgfpoint{8.30667cm}{6.15944cm}}
    \pgfusepath{clip}
    \begin{pgfscope}
      \pgfsetlinewidth{0.50pt}
      \definecolor{matfig2pgf_linecolor}{rgb}{0.000,0.000,0.000}
      \pgfsetstrokecolor{matfig2pgf_linecolor}
      \pgfsetdash{{1.50pt}{0.50pt}}{0pt}
      \pgfsetroundjoin
      \pgfplothandlerlineto
\pgfplotstreamstart
\foreach \x/\y in {1.494/1.005,1.494/1.533,1.520/1.914,1.533/2.284,1.537/2.712,1.551/3.410,1.555/4.520,1.783/4.598,1.818/4.745,1.829/4.848,1.830/5.012,1.850/5.158,1.916/5.207,1.932/5.685,1.935/5.967,1.940/6.096,2.179/6.141,2.180/6.227,2.285/6.272,2.287/6.394,2.345/6.441,2.369/6.481,2.369/6.616,2.372/6.708,2.374/6.803,2.921/6.809,2.921/6.812,3.276/6.822,3.276/6.837,3.827/6.847,3.828/6.875,3.880/6.885,3.880/6.904,3.886/6.904,3.886/6.931,3.899/6.939,3.899/6.966,3.905/6.977,3.905/7.008,3.930/7.015,3.930/7.049,3.936/7.057,3.936/7.072,5.881/7.075,5.881/7.081,5.996/7.084,5.999/7.095,6.021/7.097,6.024/7.132,6.029/7.151,6.030/7.155,11.087/7.155}
{
\pgfplotstreampoint{\pgfpoint{\x cm}{\y cm}}
}
\pgfplotstreamend
      \pgfusepath{stroke}
    \end{pgfscope}
  \end{pgfscope}
  \fi
  \makeatletter\ifpgf@draftmode\makeatother\else
  \begin{pgfscope}
    \pgfpathrectangle{\pgfpoint{1.41111cm}{1.05833cm}}{\pgfpoint{8.30667cm}{6.15944cm}}
    \pgfusepath{clip}
    \begin{pgfscope}
      \pgfsetlinewidth{0.50pt}
      \definecolor{matfig2pgf_linecolor}{rgb}{0.000,0.000,0.000}
      \pgfsetstrokecolor{matfig2pgf_linecolor}
      \pgfsetdash{}{0pt}
      \pgfsetroundjoin
      \pgfplothandlerlineto
\pgfplotstreamstart
\foreach \x/\y in {1.411/3.642,1.411/6.325,1.411/7.052,1.411/7.195,1.411/7.215,1.411/7.218,1.412/7.218,2.243/7.218,3.074/7.218,3.904/7.218,4.735/7.218,5.566/7.218,6.396/7.218,7.227/7.218,8.058/7.218,8.888/7.218,9.719/7.218}
{
\pgfplotstreampoint{\pgfpoint{\x cm}{\y cm}}
}
\pgfplotstreamend
      \pgfusepath{stroke}
    \end{pgfscope}
    \begin{pgfscope}
      \pgfsetlinewidth{0.50pt}
      \definecolor{matfig2pgf_edgecolor}{rgb}{0.000,0.000,0.000}
      \pgfsetstrokecolor{matfig2pgf_edgecolor}
      \pgfplothandlermark{\pgfpathmoveto{\pgfpoint{-3.00pt}{0pt}}\pgfpathlineto{\pgfpoint{3.00pt}{0pt}}\pgfpathmoveto{\pgfpoint{0pt}{-3.00pt}}\pgfpathlineto{\pgfpoint{0pt}{3.00pt}}\pgfpathmoveto{\pgfpoint{-2.12pt}{-2.12pt}}\pgfpathlineto{\pgfpoint{2.12pt}{2.12pt}}\pgfpathmoveto{\pgfpoint{-2.12pt}{2.12pt}}\pgfpathlineto{\pgfpoint{2.12pt}{-2.12pt}}\pgfusepath{stroke}}
\pgfplotstreamstart
\foreach \x/\y in {1.411/3.642,1.411/6.325,1.411/7.052,1.411/7.195,1.411/7.215,1.411/7.218,1.412/7.218,2.243/7.218,3.074/7.218,3.904/7.218,4.735/7.218,5.566/7.218,6.396/7.218,7.227/7.218,8.058/7.218,8.888/7.218,9.719/7.218}
{
\pgfplotstreampoint{\pgfpoint{\x cm}{\y cm}}
}
\pgfplotstreamend
    \end{pgfscope}
  \end{pgfscope}
  \fi
    \pgftext[top,x=5.55122cm,y=0.437102cm,rotate=0]{False Positive Rate $\alpha$}
    \pgftext[bottom,x=0.471982cm,y=4.11162cm,rotate=90]{True Positive Rate $\beta$}
  \begin{pgfscope}
    \definecolor{matfig2pgf_color}{rgb}{1,1,1}\pgfsetfillcolor{matfig2pgf_color}
    \pgfpathrectangle{\pgfpoint{4.3794cm}{1.18866cm}}{\pgfpoint{5.15545cm}{1.26903cm}}
    \pgfusepath{fill}
  \end{pgfscope}
  \begin{pgfscope}
    \pgfsetlinewidth{0.5pt}
    \pgfusepath{stroke}
  \end{pgfscope}
  \begin{pgfscope}
    \pgfsetlinewidth{0.5pt}
    \pgfpathrectangle{\pgfpoint{4.3794cm}{1.18866cm}}{\pgfpoint{5.15545cm}{1.26903cm}}
    \pgfusepath{stroke}
  \end{pgfscope}
    \pgftext[left,x=5.77456cm,y=1.54558cm,rotate=0]{Test Statistic $T_{\widehat{\mathit{LR}}\mathit{-FD}}$}
  \makeatletter\ifpgf@draftmode\makeatother\else
  \begin{pgfscope}
    \pgfpathrectangle{\pgfpoint{4.3794cm}{1.18866cm}}{\pgfpoint{5.15545cm}{1.26903cm}}
    \pgfusepath{clip}
    \begin{pgfscope}
      \pgfsetlinewidth{0.50pt}
      \definecolor{matfig2pgf_linecolor}{rgb}{0.000,0.000,0.000}
      \pgfsetstrokecolor{matfig2pgf_linecolor}
      \pgfsetdash{{1.50pt}{0.50pt}}{0pt}
      \pgfsetroundjoin
      \pgfplothandlerlineto
\pgfplotstreamstart
\foreach \x/\y in {4.591/1.576,5.648/1.576}
{
\pgfplotstreampoint{\pgfpoint{\x cm}{\y cm}}
}
\pgfplotstreamend
      \pgfusepath{stroke}
    \end{pgfscope}
  \end{pgfscope}
  \fi
  \makeatletter\ifpgf@draftmode\makeatother\else
  \begin{pgfscope}
    \pgfpathrectangle{\pgfpoint{4.3794cm}{1.18866cm}}{\pgfpoint{5.15545cm}{1.26903cm}}
    \pgfusepath{clip}
  \end{pgfscope}
  \fi
    \pgftext[left,x=5.77456cm,y=2.12722cm,rotate=0]{ROC Upper Bound}
  \makeatletter\ifpgf@draftmode\makeatother\else
  \begin{pgfscope}
    \pgfpathrectangle{\pgfpoint{4.3794cm}{1.18866cm}}{\pgfpoint{5.15545cm}{1.26903cm}}
    \pgfusepath{clip}
    \begin{pgfscope}
      \pgfsetlinewidth{0.50pt}
      \definecolor{matfig2pgf_linecolor}{rgb}{0.000,0.000,0.000}
      \pgfsetstrokecolor{matfig2pgf_linecolor}
      \pgfsetdash{}{0pt}
      \pgfsetroundjoin
      \pgfplothandlerlineto
\pgfplotstreamstart
\foreach \x/\y in {4.591/2.149,5.648/2.149}
{
\pgfplotstreampoint{\pgfpoint{\x cm}{\y cm}}
}
\pgfplotstreamend
      \pgfusepath{stroke}
    \end{pgfscope}
  \end{pgfscope}
  \fi
  \makeatletter\ifpgf@draftmode\makeatother\else
  \begin{pgfscope}
    \pgfpathrectangle{\pgfpoint{4.3794cm}{1.18866cm}}{\pgfpoint{5.15545cm}{1.26903cm}}
    \pgfusepath{clip}
    \begin{pgfscope}
      \pgfsetlinewidth{0.50pt}
      \definecolor{matfig2pgf_edgecolor}{rgb}{0.000,0.000,0.000}
      \pgfsetstrokecolor{matfig2pgf_edgecolor}
      \pgfplothandlermark{\pgfpathmoveto{\pgfpoint{-3.00pt}{0pt}}\pgfpathlineto{\pgfpoint{3.00pt}{0pt}}\pgfpathmoveto{\pgfpoint{0pt}{-3.00pt}}\pgfpathlineto{\pgfpoint{0pt}{3.00pt}}\pgfpathmoveto{\pgfpoint{-2.12pt}{-2.12pt}}\pgfpathlineto{\pgfpoint{2.12pt}{2.12pt}}\pgfpathmoveto{\pgfpoint{-2.12pt}{2.12pt}}\pgfpathlineto{\pgfpoint{2.12pt}{-2.12pt}}\pgfusepath{stroke}}
\pgfplotstreamstart
\foreach \x/\y in {5.120/2.149}
{
\pgfplotstreampoint{\pgfpoint{\x cm}{\y cm}}
}
\pgfplotstreamend
    \end{pgfscope}
  \end{pgfscope}
  \fi
  \makeatletter\ifpgf@draftmode\makeatother\pgftext[x=5cm,y=3.75cm]{\Huge{DRAFT}}\fi
\end{tikzpicture}}
    \caption{ROC curve of $T_{\widehat{\mathit{LR}}\mathit{-FD}}$ using fixed-degree models for both the null and alternate hypotheses. (The stepped appearance of the curve is an artifact of the importance sampling weights.)  Also shown is an ROC upper bound, obtained using knowledge of the true group membership for each node}
    \label{fig:FD_ROC}
\end{figure}
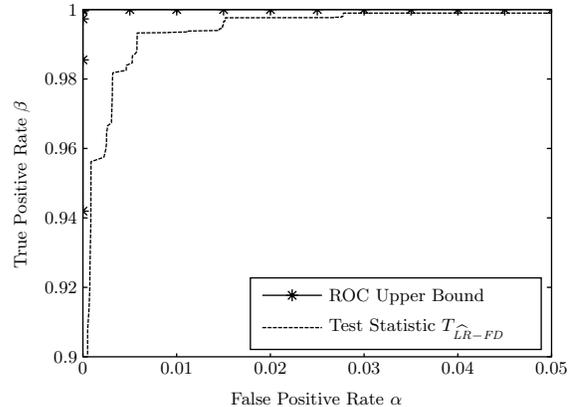
the estimated p-value of this data set remains below $10^{-3}$.
\end{examp}

Note that specification of parameters $p_{00}$, $p_{01}$, and $p_{11}$ was required to generate Fig.~\ref{fig:FD_ROC} via simulation; here, we manually fit these three parameters to the data, starting with their estimates under the two-group stochastic block model, until the likelihood of the observed data approached the median likelihood under our parameterization.  A more formal fitting procedure could of course be adopted in practice.

\section{$\!\!$Open$\!$ Problems$\!$ in$\!$ Network$\!$ Inference$\!$}\label{sec:disc}

The examples of Sections~\ref{sec:modelSpec} and~\ref{sec:DataExample} were designed to be illustrative, and yet they also serve to illuminate broader questions that arise as we seek to extend classical notions of statistics to network data. As we have seen in Section~\ref{sec:modelSpec}, for instance, the inclusion of latent $k$-ary categorical covariates immediately necessitates a variety of combinatorial calculations.  The increasing prevalence of large, complex network data sets presents an even more significant computational challenge for statistical inference.  Indeed, longstanding inferential frameworks---as exemplified by the hypothesis tests of Section~\ref{sec:DataExample}, for instance---are crucial to the analysis of networks and relational data, and yet their implementations can prove remarkably difficult even for small data sets.

To address these broader questions and impact the future of network inference, we believe that statisticians should focus on the following three main categories of open problems, whose descriptions comprise this remainder of this section:
\begin{enumerate}
\item We must work to specify models that can more realistically describe observed network data.  For instance, the fixed-degree models introduced earlier account explicitly for heterogeneous degree sequences;  in the case of large-scale network data sets, even more flexible models are needed.\vspace{0.5\baselineskip}%
\item We must build approximations to these models for which likelihood maximization can be readily achieved, along with tools to evaluate the quality of these approximations.  The spectral partitioning approach featured in our examples of Section~\ref{sec:DataExample} serves as a prime example; however, validation of approximate inference procedures remains an important open question.\vspace{0.5\baselineskip}%
\item We must seek to understand precisely how network sampling influences our statistical analyses.  In addition to better accounting for data gathering mechanisms, sampling can serve as a method of data reduction.  This in turn will enable the application of a variety of methods to data sets much larger than those exhibited here.
\end{enumerate}

\subsection{Model Elicitation and Selection}

More realistic network models can only serve to benefit statistical inference, regardless of their computational or mathematical convenience~\citep{banks1998mmr}.  Models tailored to different fields, and based on theory fundamental to specific application areas, are of course the long-term goal---with the exponential random graph models reviewed by~\citet{anderson99:_a_p_star_primer} and~\citet{snijders06:_new_specifications_for_ERGMs} among the most successful and widely known to date.  However, additional work to determine more general models for network structure will also serve to benefit researchers and practitioners alike.  As detailed in the Appendix, there are presently several competing models of this type, each with its own merits: stochastic block models~\citep{wang87:_stochastic_blockmodels_for_directed_graphs}, block models with mixed membership~\citep{airoldi07:_combining_stochastic_block_models}, and structural models that explicitly incorporate information regarding the degree sequence in addition to group membership~\citep{chung03:_spectra_of_random_graphs_with_given_expected_degrees}.

At present, researchers lack a clearly articulated strategy for selecting between these different modeling approaches---the goodness-of-fit procedures
of~\citet{hunter08:_goodness_of_fit_of_social_network_models}, based on graphical comparisons of various network statistics, provide a starting point, but comparing the complexity of these different modeling strategies poses a challenge.  Indeed, it is not even entirely clear how best to select the number of groups used in a single modeling strategy alone.  For the data of Section~\ref{ex:Zachary}, for example, we restricted our definition of network structure to be a binary division of the data into two groups, whereas many observed data sets may cluster into an \emph{a priori} unknown number of groups.

It is also worth noting that many different fields of mathematics may provide a source for network data models.  While graph theory forms a natural starting point, other approaches based on a combination of random matrices, algebra, and geometry may also prove useful.  For example, the many graph partitioning algorithms based on spectral methods suggest the use of corresponding generative models based on the eigenvectors and eigenvalues of the graph Laplacian.  The primary challenge in this case appears to be connecting such models to the observed data matrix $\bm{A}$, which typically consists of binary entries.

\subsection{Approximate Inference and Validation}

Computationally or mathematically convenient models will also continue to play a key role in network analysis.  Even simple generic models of structure are very high-dimensional, and with network data sets commonly consisting of thousands to millions of nodes, model dimensionality spirals out of control at an impossible rate.  Somehow this fundamental challenge of network data---how to grapple with the sheer number of relational observations---must be turned into a strength so that further analysis may proceed.  Reducing the dimensionality through an approximate clustering is an excellent first step to build upon, but computationally realizable inference schemes must also follow in turn.

The usefulness of such approximations will ultimately be determined by the extent to which evaluation tools can be developed for massive data sets.  Whenever models are sufficiently complex to necessitate approximate inference procedures, such models must be paired with mechanisms to relate the quality of the resulting analysis back to the original problem and model specification.  Indeed, assurances are needed to convince thoughtful practitioners that analyzing a different model, or maximizing a quantity other than desired likelihood, is a useful exercise.

Other approaches to validation may focus on the outcome of the analysis in some way, rather than its theoretical underpinnings.  With ground truth by its very definition available only for small-scale illustrative problems, or for those which are  generally felt to have already been solved, prediction may provide a valuable substitute.  By monitoring the results of approximation over time relative to revealed truth, confidence in the adopted inference procedure may be grown.

\subsection{Sampling, Missingness, and Data Reduction}

A final concern is to better understand how sampling mechanisms influence network inference.  Consider that
two critical assumptions almost always underpin the vast majority of contemporary network analyses: First, that all links within the collection of observed nodes have been accounted for; and second, that observed nodes within the network comprise the only nodes of interest.  In general, neither of these assumptions may hold in practice.

To better understand the pitfalls of the first assumption, consider that while observing the presence of a link between nodes is typically a feasible and well defined task, observing the \emph{absence} of a link can in many cases pose a substantial challenge.  Indeed, careful reflection often reveals that zero entries in relational data sets are often better thought of as unobserved~\citep{clauset2008hsa, marchette2008pul}.  The implications of this fact for subsequent analysis procedures---as well as on approximate likelihood maximization procedures and spectral methods in particular---remains unclear.

The second assumption, that all nodes of interest have in fact been recorded, also appears rarely justified in practice.  Indeed, it seems an artifact of this assumption that most commonly studied data sets consist of nodes which form a connected network.  While in some cases the actual network may in fact consist of a single connected component, researchers may have unwittingly selected their data conditioned upon its appearance in the largest connected component of a much larger network.  How this selection in turn may bias the subsequent fitting of models has only recently begun to be investigated~\citep{handcock2009msn}.

A better understanding of missingness may also lend insight into optimal sampling procedures.  Although researchers themselves may lack influence over data gathering mechanisms, the potential of such methods for data reduction is clear.  One particularly appealing approach is to first sample very large network data sets in a controlled manner, and then apply exact analysis techniques.  In some cases the resultant approximation error can be bounded~\citep{belabbas08:_spect_method_machin_learn}, implying that the effects on inferential procedures in question can be quantified.

Other data reduction techniques may also help to meet the computational challenges of network analysis;  for example, \citet{krishnamurthy07:_sampling_large_internet_topologies} examined contractions of nodes into groups as a means of lessening data volume.  Such strategies of reducing network size while preserving relevant information provide an alternative to approximate likelihood maximization that is deserving of further study.

\section{Conclusion}

In many respects, the questions being asked of network data sets are not at all new to statisticians.  However, the increasing prevalence of large networks in contemporary application areas gives rise to both challenges and opportunities for statistical science. Tests for detecting network structure in turn form a key first step toward more sophisticated inferential procedures, and moreover provide practitioners with much-needed means of formal data analysis.

Classical inferential frameworks are precisely what is most needed in practice, and yet as we have seen, their exact implementation can prove remarkably difficult in the setting of modern high-dimensional, non-Euclidean network data. To this end, we hope that this paper has succeeded in helping to chart a path toward the ultimate goal of a unified and coherent framework for the statistical analysis of large-scale network data sets.

\section*{Acknowledgments}

Research supported in part by the National Science Foundation under Grants~DMS-0631636 and CBET-0730389, and the Defense Advanced Research Projects Agency under Grant No.~HR0011-07-1-0007.

\appendix

\section*{Appendix: A Review of Approaches to Network Data Analysis}

Three canonical problems in network data analysis have consistently drawn attention across different contexts: network model elicitation, network model inference, and methods of approximate inference.

\subsection{Model Elicitation}
\label{sec:modelElic}

With new network data sets being generated or discovered at rapid rates in a wide variety of fields, model elicitation---independent even of model selection---remains an important topic of investigation. Although graph theory provides a natural starting point for identifying possible models for graph-valued data, practitioners have consistently found that models such as Erd\"{o}s-R\'{e}nyi lack sufficient explanatory power for complex data sets.  Its inability to model all but the simplest of degree distributions has forced researchers to seek out more complicated models.

\citet{barabasi02:_linked} and \citet{palla05:_uncovering_the_overlapping_community_structure_of_networks} survey a wide variety of network data sets and conclude that commonly encountered degree sequences follow a power law or similarly heavy-tailed distribution; the Erd\"{o}s-R\'{e}nyi model, with its marginally binomial degree distribution, is obviously insufficient to describe such data sets. \citet{barabasi99:_emergence_of_scaling_in_random_networks} introduced an alternative by way of a generative network modeling scheme termed ``preferential attachment'' to explicitly describe power-law degree sequences. Under this scheme, nodes are added sequentially to the graph, being preferentially linked to existing nodes based on the current degree sequence.  A moment's reflection will convince the reader that this model is in fact an example of a Dirichlet process~\citep{pemantle07:_survey_of_random_processes_with_reinforcement}.

Though the preferential attachment approach serves to describe the observed degree sequences of many networks, it can fall short of correctly modeling their patterns of connectivity~\citep{li05:_towards_a_theory_of_scale_free_graphs}; moreover, heterogenous degree sequences many not necessarily follow power laws.  A natural solution to both problems is to condition on the observed degree sequence as in Section~\ref{sec:_two_fixed_degree_models} and consider the connections between nodes to be random.  As described earlier, the difficulties associated with simulating fixed-degree simple graphs have historically dissuaded researchers from this direction, and hence fixed-degree models have not yet seen wide use in practice.

As an alternative to fixed-degree models, researchers have instead focused on the so-called configuration
model~\citep{newman01:_random_graphs_with_arbitrary_degree_distributions} as well as models which yield graphs of given expected degree~\citep{chung03:_spectra_of_random_graphs_with_given_expected_degrees}.  The configuration model specifies the degree sequence exactly, as with the case of fixed-degree models, but allows both multiple links between nodes and ``self-loops'' in order to gain a simplified asymptotic analysis.  Models featuring given expected degrees specify only the expected degree of each node---typically set equal to the observed degree---and allow the degree sequence to be random.  Direct simulation becomes possible if self-loops and multiple links are allowed, thus enabling approximate inference methods of the type described in Section~\ref{sec:ApproximateInference}. However, observed network data sets do not always exhibit either of these phenomena, thus rendering the inferential utility of these models highly dependent on context.  In the case of very large data sets, for example, the possible presence or absence of multiple connections or self-loops in the model may be irrelevant to describing the data on a coarse scale. When it becomes necessary to model network data at a fine scale, however, a model which allows for these may be insufficiently realistic.

Graph models may equally well be tailored to specific fields.  For example, sociologists and statisticians working in concert have developed a class of well-known approaches collectively known as exponential random graph models (ERGMs) or alternatively as $p^*$ models.  Within this class of models, the probability of nodes being linked to each other depends explicitly on parameters that control well-defined sufficient statistics; practitioners draw on sociological theory to determine which connectivity statistics are critical to include within the model.  A key advantage of these models is that they can readily incorporate covariates into their treatment of connectivity properties.  For a detailed review, along with a discussion of some of the latest developments in the field of social networks, the reader is referred to~\citet{anderson99:_a_p_star_primer} and~\citet{snijders06:_new_specifications_for_ERGMs}.

Since their original introduction, ERGMs have been widely adopted as models for social networks. They have not yet, however, been embraced to the same extent by researchers outside of social network analysis.  Sociologists can rely on existing theory to select models for how humans form relationships with each other; researchers in other fields, though, often cannot appeal to equivalent theories.  For exploratory analysis, they may require more generic models to describe their data, appearing to prefer models with a latent vector of covariates to capture probabilistically exchangeable blocks.  Indeed, as noted in Section~\ref{sec:E-R}, this approach falls under
the general category of stochastic block modeling. \citet{wang87:_stochastic_blockmodels_for_directed_graphs} detail similarities and differences between this approach and the original specification of ERGMs.

Stochastic block modeling, though relatively generic, may still fail to adequately describe networks in which nodes roughly group together, yet in large part fail to separate into distinct clusters.  In cases such as this, where stochastic exchangeability is too strong an assumption, standard block modeling breaks down.  To this end, two possible modeling solutions have been explored to date in the literature. \citet{hoff02:_latent_space_approaches_to_sna} introduced a latent space approach, describing the probability of
connection as a function of distances between nodes in an unobserved space of known dimensionality. In this model, the observed grouping of nodes is a result of their proximity in this latent space. In contrast,
\citet{airoldi07:_combining_stochastic_block_models} retained the
explicit grouping structure that stochastic block modeling provides, but introduced the idea of mixed group membership to describe nodes that fall between groups.  Node membership here is a vector describing partial membership in all groups, rather than an indicator variable specifying a single group membership.

\subsection{Model Fitting and Inference}

Even when a model or class of models for network data can be specified, realizing inference can be challenging.  One of the oldest uses of random graph models is as a null; predating the computer, \citet{moreno38:_statistics_of_social_configurations} simulated a random graph model quite literally by hand in order to tabulate null model statistics.  These authors drew cards out of a ballot shuffling apparatus to generate graphs of the same size as a social network of schoolgirls they had observed.  Comparing the observed statistics to the distribution of tabulated statistics, they rejected the hypothesis that the friendships they were observing were formed strictly by chance.

Asymptotic tests may alleviate the need for simulation in cases of large network data sets, and are available for  certain models and test statistics---the $\chi^2$-test of Section~\ref{sec:NetworkStrutureDetection} being one such example.  As another example, \citet{holland81:_exponential_family_for_directed_graphs} developed asymptotic tests based on likelihood ratio statistics to select between different ERGMs.  Sociologists and statisticians
together have developed results for other test statistics as well, many of which are reviewed
by~\citet{wasserman94:_social_network_analysis}.

A desire upon the rejection of a null model, of course, is the fitting of an alternate.  However, as demonstrated in Section~\ref{sec:ApproximateInference}, direct fitting by maximum likelihood can prove computationally costly, even for basic network models.  A common solution to maximizing the likelihood under an ERGM, for example, is to employ a Markov chain Monte Carlo
strategy~\citep{snijders06:_new_specifications_for_ERGMs}.  \citet{handcock07:_model_based_clustering_for_social_networks}
also used such methods to maximize the likelihood of a latent space network model; additionally, these authors suggested a faster, though approximate, two-stage maximization routine.

Other researchers have employed greedy algorithms to maximize the model likelihood.  \citet{newman07:_mixture_models_and_exploratory_analysis} used expectation-maximization (EM) to fit a network model related to stochastic block modeling.  Relaxing the precise requirements of the EM algorithm, both
\citet{hofman08:_bayesian_approach_to_network_modularity} and~\citet{airoldi08:_mixed_membership_stochastic_blockmodels} have applied a variational Bayes approach
(see, e.g., \citet{jordan99:_introduction_to_variational_methods}) to find maximum likelihood estimates of parameters under a stochastic block model.  \citet{reichardt04:_detecting_fuzzy_community_structure} applied
simulated annealing to maximize the likelihood of network data under a Potts model, a generalization of the Ising model.  \citet{rosvall07:_an_information_theoretic_framework_for_resolving_community_structure,
rosvall08:_maps_of_random_walks_on_complex_networks} have also employed simulated annealing in network inference  in order to maximize information-theoretic functionals of the data.

Following any kind of model fitting procedure, a goodness-of-fit test of some kind is clearly desirable.  Yet, researchers have thus far struggled to find a clear solution to this problem.  \citet{hunter08:_goodness_of_fit_of_social_network_models} have proposed a general method of accumulating a wide set of network statistics, and comparing them graphically to the distribution of these same statistics under a fitted model.  Networks which fit well should in turn exhibit few statistics that deviate far from those simulated from the corresponding model.

\subsection{Approximate Inference Procedures}

In most cases of practical interest, and in particular for large network data sets, model likelihoods cannot be maximized in a computationally feasible manner, and researchers must appeal to a heuristic that yields some approximately maximized quantity.  With this goal in mind, the idea of likelihood maximization has been subsumed by the idea of fast graph partitioning described in Section~\ref{sec:ApproximateInference}, as it is the process of
determining group membership which typically poses the most computational challenges.   The invention of new algorithms that can quickly partition large graphs is clearly of great utility here.

\subsubsection{Algorithmic Approaches}

Computer scientists and physicists have long been active in the creation of new graph partitioning algorithms.  In addition to techniques such as spectral bisection, many researchers have also noted that the inherently sparse nature of most real-world adjacency structures enables faster implementations of spectral methods (see, e.g.,~\citet{white05:_spectral_clustering_approach_to_finding_communities}).

Researchers have sought to also incorporate graph partitioning concepts that allow for multiple partitions of varying sizes.  Some researchers, such as \citet{eckmann02:_curvature_of_colinks_uncovers_layers} and \citet{radicchi04:_defining_and_identifying_communities_in_networks}, have attempted to use strictly local statistics to aid in the clustering of nodes into multiple partitions.  \citet{girvan02:_community_structure_in_networks} focused in contrast on global statistics, by way of measures of the centrality of a node relative
to the rest of the graph.  This line of reasoning eventually resulted in the introduction of modularity~\citep{newman06:_modularity_and_community_structure} as a global statistic to relate the observed number
of edges between groups to their expected number under the configuration model outlined in Section~\ref{sec:modelElic} above.  Spectral clustering methods can also be applied to the task of approximately maximizing modularity, in a manner that enables both group size and number to vary.  A wide variety of alternative maximization approaches have been applied as well: Both \citet{wang07:_detecting_community_structure_by_optimal_rearrangement_clustering}
and \citet{brandes08:_on_modularity_clustering} review the computational difficulties associated with maximization of the modularity statistic, and relate this to known combinatorial optimization problems.  \citet{fortunato07:_community_structure_in_graphs} review many recently proposed maximization routines and contrast them with traditional methods.

\subsubsection{Evaluation of Efficacy}

Approximate procedures in turn require some way to evaluate the departure from exact likelihood maximization.  Thus far, a clear way to evaluate partitions found through the various heuristics cited above has not yet emerged, though many different approaches have been proposed.  Both~\citet{massen06:_thermodynamics_of_community_structure}
and~\citet{karrer08:_robustness_of_community_structure} have explored ways to test the statistical significance of the output of graph partitioning algorithms.  Their methods attempt to determine whether a model which lacks structure could equally well explain the group structure inferred from the data. These approaches, though
distinct from one another, are both akin to performing a permutation test---a method known to be effective when applied to more general cases of clustering.  \citet{carley1993nin} apply this exact idea to test for structure when group memberships are given.

Other researchers have attempted a more empirical approach to the problem of partition evaluation by adopting a metric to measure the distance between found and ``true'' partitions.  Such distances are then examined for a variety of data sets and simulated cases for which the true partition is assumed known.  In this vein \citet{danon05:_comparing_community_structure_identification} specified an explicit probability model for structure and compared how well different graph partitioning schemes recovered the true subgroups of data, ranking them by both execution time as well as average distance between true and found partitions.  \citet{gustafsson06:_comparison_and_validation_of_community_structures}  performed a similar comparison, along with a study of differences in ``found'' partitions between algorithms for
several well-known data sets, including the karate club data of Section~\ref{ex:Zachary}.  They found that standard clustering algorithms (e.g., $k$-means) sometimes outperform more specialized network partition algorithms.  Finally, \citet{fortunato07:_resolution_limit_in_community_detection} have
undertaken theoretical investigations of the sensitivity and power of a particular partitioning algorithm to detect subgroups below a certain size.

\bibliographystyle{chicago} 

\begin{thebibliography}{}

\bibitem[\protect\citeauthoryear{Adamic and Huberman}{Adamic and
  Huberman}{2000}]{adamic00:_power_law_distribution_of_www}
Adamic, L.~A. and B.~A. Huberman (2000).
\newblock Power-law distribution of the {W}orld {W}ide {W}eb.
\newblock {\em Science\/}~{\em 287}, 2115.

\bibitem[\protect\citeauthoryear{Airoldi, Blei, Fienberg, and Xing}{Airoldi
  et~al.}{2007}]{airoldi07:_combining_stochastic_block_models}
Airoldi, E.~M., D.~M. Blei, S.~E. Fienberg, and E.~P. Xing (2007).
\newblock Combining stochastic block models and mixed membership for
  statistical network analysis.
\newblock In E.~M. Airoldi, D.~M. Blei, S.~E. Fienberg, A.~Goldenberg, E.~P.
  Xing, and A.~X. Zheng (Eds.), {\em Papers from the ICML 2006 Workshop on
  Statistical Network Analysis}, pp.\  57--74. Berlin: Springer.

\bibitem[\protect\citeauthoryear{Airoldi, Blei, Fienberg, and Xing}{Airoldi
  et~al.}{2008}]{airoldi08:_mixed_membership_stochastic_blockmodels}
Airoldi, E.~M., D.~M. Blei, S.~E. Fienberg, and E.~P. Xing (2008).
\newblock Mixed membership stochastic blockmodels.
\newblock {\em J. Machine Learn. Res.\/}~{\em 9}, 1981--2014.

\bibitem[\protect\citeauthoryear{Altman, Lausen, Sauerbrei, and
  Schumacher}{Altman et~al.}{1994}]{altman94:_dangers_using_optimal_cutpoints}
Altman, D.~G., B.~Lausen, W.~Sauerbrei, and M.~Schumacher (1994).
\newblock Dangers of using ``optimal'' cutpoints in the evaluation of
  prognostic factors.
\newblock {\em J. Natl. Cancer Inst.\/}~{\em 86}, 829--835.

\bibitem[\protect\citeauthoryear{Anderson, Wasserman, and Crouch}{Anderson
  et~al.}{1999}]{anderson99:_a_p_star_primer}
Anderson, C.~J., S.~Wasserman, and B.~Crouch (1999).
\newblock A $p^*$ primer: Logit models for social networks.
\newblock {\em Social Networks\/}~{\em 21}, 37--66.

\bibitem[\protect\citeauthoryear{Banks and Constantine}{Banks and
  Constantine}{1998}]{banks1998mmr}
Banks, D. and G.~M. Constantine (1998).
\newblock Metric models for random graphs.
\newblock {\em J. Classificat.\/}~{\em 15}, 199--223.

\bibitem[\protect\citeauthoryear{Barab{\'a}si}{Barab{\'a}si}{2002}]{barabasi02%
:_linked}
Barab{\'a}si, A. (2002).
\newblock {\em Linked: The new science of networks}.
\newblock Cambridge, MA: Perseus Publishing.

\bibitem[\protect\citeauthoryear{Barab{\'a}si and Albert}{Barab{\'a}si and
  Albert}{1999}]{barabasi99:_emergence_of_scaling_in_random_networks}
Barab{\'a}si, A. and R.~Albert (1999).
\newblock Emergence of scaling in random networks.
\newblock {\em Science\/}~{\em 286}, 509--512.

\bibitem[\protect\citeauthoryear{Batada, Reguly, Breitkreutz, Boucher,
  Breitkreutz, Hurst, and Tyers}{Batada
  et~al.}{2006}]{batada06:_view_of_the_yeast_PPI_network}
Batada, N.~N., T.~Reguly, A.~Breitkreutz, L.~Boucher, B.-J. Breitkreutz, L.~D.
  Hurst, and M.~Tyers (2006).
\newblock Stratus not altocumulus: {A} new view of the yeast protein
  interaction network.
\newblock {\em PLoS Biology\/}~{\em 4}, 1720--1731.

\bibitem[\protect\citeauthoryear{Belabbas and Wolfe}{Belabbas and
  Wolfe}{2009}]{belabbas08:_spect_method_machin_learn}
Belabbas, M.-A. and P.~J. Wolfe (2009).
\newblock Spectral methods in machine learning and new strategies for very
  large data sets.
\newblock {\em Proc. Natl. Acad. Sci. USA\/}~{\em 106}, 369--374.

\bibitem[\protect\citeauthoryear{Blitzstein and Diaconis}{Blitzstein and
  Diaconis}{2006}]{blitzstein06:_a_sequential_importance_sampling_algorithm_fo%
r_random_graphs}
Blitzstein, J. and P.~Diaconis (2006).
\newblock A sequential importance sampling algorithm for generating random
  graphs with prescribed degrees.
\newblock Unpublished manuscript, available online.

\bibitem[\protect\citeauthoryear{Bollob{\'a}s and Scott}{Bollob{\'a}s and
  Scott}{2004}]{bollobas2004mcr}
Bollob{\'a}s, B. and A.~D. Scott (2004).
\newblock {M}ax {C}ut for random graphs with a planted partition.
\newblock {\em Combinat. Probab. Comput.\/}~{\em 13}, 451--474.

\bibitem[\protect\citeauthoryear{Brandes, Delling, Gaertler, G\"orke, Hoefer,
  Nikoloski, and Wagner}{Brandes
  et~al.}{2008}]{brandes08:_on_modularity_clustering}
Brandes, U., D.~Delling, M.~Gaertler, R.~G\"orke, M.~Hoefer, Z.~Nikoloski, and
  D.~Wagner (2008).
\newblock On modularity clustering.
\newblock {\em IEEE Trans. Knowl. Data Eng.\/}~{\em 20}, 172--188.

\bibitem[\protect\citeauthoryear{Carley and Banks}{Carley and
  Banks}{1993}]{carley1993nin}
Carley, K. and D.~Banks (1993).
\newblock Nonparametric inference for network data.
\newblock {\em J. Math. Sociol.\/}~{\em 18}, 1--26.

\bibitem[\protect\citeauthoryear{Ceyhan, Priebe, and Marchette}{Ceyhan
  et~al.}{2007}]{ceyhan2007nfr}
Ceyhan, E., C.~E. Priebe, and D.~J. Marchette (2007).
\newblock A new family of random graphs for testing spatial segregation.
\newblock {\em Canad. J. Statist.\/}~{\em 35}, 27--50.

\bibitem[\protect\citeauthoryear{Chung, Lu, and Vu}{Chung
  et~al.}{2003}]{chung03:_spectra_of_random_graphs_with_given_expected_degrees}
Chung, F., L.~Lu, and V.~Vu (2003).
\newblock Spectra of random graphs with given expected degrees.
\newblock {\em Proc. Natl. Acad. Sci. USA\/}~{\em 100}, 6313--6318.

\bibitem[\protect\citeauthoryear{Chung}{Chung}{1997}]{chung1997sgt}
Chung, F. R.~K. (1997).
\newblock {\em Spectral Graph Theory}.
\newblock Providence, RI: American Mathematical Society.

\bibitem[\protect\citeauthoryear{Clauset, Moore, and Newman}{Clauset
  et~al.}{2008}]{clauset2008hsa}
Clauset, A., C.~Moore, and M.~E.~J. Newman (2008).
\newblock Hierarchical structure and the prediction of missing links in
  networks.
\newblock {\em Nature\/}~{\em 453}, 98--101.

\bibitem[\protect\citeauthoryear{Danon, Diaz-Guilera, Duch, and Arenas}{Danon
  et~al.}{2005}]{danon05:_comparing_community_structure_identification}
Danon, L., A.~Diaz-Guilera, J.~Duch, and A.~Arenas (2005).
\newblock Comparing community structure identification.
\newblock {\em J Statist. Mech.\/}~{\em 9}, P09008.

\bibitem[\protect\citeauthoryear{de~Solla~Price}{de~Solla~Price}{1965}]{desoll%
aprice65:_networks_of_scientific_papers}
de~Solla~Price, D.~J. (1965).
\newblock Networks of scientific papers.
\newblock {\em Science\/}~{\em 149}, 510--515.

\bibitem[\protect\citeauthoryear{Dunne, Williams, and Martinez}{Dunne
  et~al.}{2002}]{dunne02:_food_web_structure}
Dunne, J.~A., R.~J. Williams, and N.~D. Martinez (2002).
\newblock Food-web structure and network theory: {T}he role of connectance and
  size.
\newblock {\em Proc. Natl. Acad. Sci. USA\/}~{\em 99}, 12917--12922.

\bibitem[\protect\citeauthoryear{Eagle, Pentland, and Lazer}{Eagle
  et~al.}{2008}]{eagle07:_inferring_social_network_structure_using_mobile_phon%
e_data}
Eagle, N., A.~Pentland, and D.~Lazer (2008).
\newblock Mobile phone data for inferring social network structure.
\newblock In H.~Liu, J.~J. Salerno, and M.~J. Young (Eds.), {\em Social
  Computing, Behavioral Modeling, and Prediction}, pp.\  79--88. New York:
  Springer.

\bibitem[\protect\citeauthoryear{Eckmann and Moses}{Eckmann and
  Moses}{2002}]{eckmann02:_curvature_of_colinks_uncovers_layers}
Eckmann, J.-P. and E.~Moses (2002).
\newblock Curvature of co-links uncovers hidden thematic layers in the {W}orld
  {W}ide {W}eb.
\newblock {\em Proc. Natl. Acad. Sci. USA\/}~{\em 99}, 5825--5829.

\bibitem[\protect\citeauthoryear{Erd{\"{o}}s and R{\'e}nyi}{Erd{\"{o}}s and
  R{\'e}nyi}{1959}]{erdos59:_on_random_graphs}
Erd{\"{o}}s, P. and A.~R{\'e}nyi (1959).
\newblock On random graphs.
\newblock {\em Publicat. Mathemat.\/}~{\em 6}, 290--297.

\bibitem[\protect\citeauthoryear{Fiedler}{Fiedler}{1973}]{fiedler73:_algebraic%
_connectivity_of_graphs}
Fiedler, M. (1973).
\newblock Algebraic connectivity of graphs.
\newblock {\em Czech. Math. J.\/}~{\em 23}, 298--305.

\bibitem[\protect\citeauthoryear{Fortunato and Barth\'elemy}{Fortunato and
  Barth\'elemy}{2007}]{fortunato07:_resolution_limit_in_community_detection}
Fortunato, S. and M.~Barth\'elemy (2007).
\newblock Resolution limit in community detection.
\newblock {\em Proc. Natl. Acad. Sci. USA\/}~{\em 104}, 36--41.

\bibitem[\protect\citeauthoryear{Fortunato and Castellano}{Fortunato and
  Castellano}{2007}]{fortunato07:_community_structure_in_graphs}
Fortunato, S. and C.~Castellano (2007).
\newblock Community structure in graphs.
\newblock Unpublished manuscript, available online.

\bibitem[\protect\citeauthoryear{Gilbert}{Gilbert}{1959}]{gilbert59:_random_gr%
aphs}
Gilbert, E.~N. (1959).
\newblock Random graphs.
\newblock {\em Ann. Math. Stat.\/}~{\em 30}, 1141--1144.

\bibitem[\protect\citeauthoryear{Girvan and Newman}{Girvan and
  Newman}{2002}]{girvan02:_community_structure_in_networks}
Girvan, M. and M.~E.~J. Newman (2002).
\newblock Community structure in social and biological networks.
\newblock {\em Proc. Natl. Acad. Sci. USA\/}~{\em 99}, 7821--7826.

\bibitem[\protect\citeauthoryear{Gustafsson, H{\"o}rnquist, and
  Lombardi}{Gustafsson
  et~al.}{2006}]{gustafsson06:_comparison_and_validation_of_community_structur%
es}
Gustafsson, M., M.~H{\"o}rnquist, and A.~Lombardi (2006).
\newblock Comparison and validation of community structures in complex
  networks.
\newblock {\em Physica A: Statist Mech. Appl.\/}~{\em 367}, 559--576.

\bibitem[\protect\citeauthoryear{Handcock and Gile}{Handcock and
  Gile}{2009}]{handcock2009msn}
Handcock, M.~S. and K.~J. Gile (2009).
\newblock Modeling social networks from sampled data.
\newblock {\em Ann. Appl. Statist\/}.
\newblock In press.

\bibitem[\protect\citeauthoryear{Handcock, Raftery, and Tantrum}{Handcock
  et~al.}{2007}]{handcock07:_model_based_clustering_for_social_networks}
Handcock, M.~S., A.~E. Raftery, and J.~M. Tantrum (2007).
\newblock Model-based clustering for social networks.
\newblock {\em J. Roy. Statist. Soc. A\/}~{\em 170}, 301--354.

\bibitem[\protect\citeauthoryear{Hoff, Raftery, and Handcock}{Hoff
  et~al.}{2002}]{hoff02:_latent_space_approaches_to_sna}
Hoff, P.~D., A.~E. Raftery, and M.~S. Handcock (2002).
\newblock Latent space approaches to social network analysis.
\newblock {\em J. Am. Statist. Ass.\/}~{\em 97}, 1090--1098.

\bibitem[\protect\citeauthoryear{Hofman and Wiggins}{Hofman and
  Wiggins}{2008}]{hofman08:_bayesian_approach_to_network_modularity}
Hofman, J.~M. and C.~H. Wiggins (2008).
\newblock Bayesian approach to network modularity.
\newblock {\em Phys. Rev. Lett.\/}~{\em 100}, 258701(1--4).

\bibitem[\protect\citeauthoryear{Holland and Leinhardt}{Holland and
  Leinhardt}{1981}]{holland81:_exponential_family_for_directed_graphs}
Holland, P.~W. and S.~Leinhardt (1981).
\newblock An exponential family of probability distributions for directed
  graphs.
\newblock {\em J. Am. Statist. Ass.\/}~{\em 76}, 33--50.

\bibitem[\protect\citeauthoryear{Hunter, Goodreau, and Handcock}{Hunter
  et~al.}{2008}]{hunter08:_goodness_of_fit_of_social_network_models}
Hunter, D.~R., S.~M. Goodreau, and M.~S. Handcock (2008).
\newblock Goodness of fit of social network models.
\newblock {\em J. Am. Statist. Ass.\/}~{\em 103}, 248--258.

\bibitem[\protect\citeauthoryear{Jin, Parkes, and Wolfe}{Jin
  et~al.}{2007}]{jin07:_analysis_of_bidding_networks}
Jin, R. K.-X., D.~C. Parkes, and P.~J. Wolfe (2007).
\newblock Analysis of bidding networks in {eB}ay: {A}ggregate preference
  identification through community detection.
\newblock In C.~Geib and D.~Pynadath (Eds.), {\em Plan, Activity, and Intent
  Recognition (PAIR): Papers from the 2007 AAAI Workshop}, pp.\  66--73. Menlo
  Park, CA: AAAI Press.

\bibitem[\protect\citeauthoryear{Jordan, Ghahramani, Jaakkola, and Saul}{Jordan
  et~al.}{1999}]{jordan99:_introduction_to_variational_methods}
Jordan, M.~I., Z.~Ghahramani, T.~S. Jaakkola, and L.~K. Saul (1999).
\newblock An introduction to variational methods for graphical models.
\newblock {\em Machine Learn.\/}~{\em 37}, 183--233.

\bibitem[\protect\citeauthoryear{Karrer, Levina, and Newman}{Karrer
  et~al.}{2008}]{karrer08:_robustness_of_community_structure}
Karrer, B., E.~Levina, and M.~E.~J. Newman (2008).
\newblock Robustness of community structure in networks.
\newblock {\em Phys. Rev. E\/}~{\em 77}, 46119(1--9).

\bibitem[\protect\citeauthoryear{Krishnamurthy, Faloutsos, Chrobak, Cui, Lao,
  and Percus}{Krishnamurthy
  et~al.}{2007}]{krishnamurthy07:_sampling_large_internet_topologies}
Krishnamurthy, V., M.~Faloutsos, M.~Chrobak, J.~H. Cui, L.~Lao, and A.~G.
  Percus (2007).
\newblock Sampling large {I}nternet topologies for simulation purposes.
\newblock {\em Comput. Networks\/}~{\em 51}, 4284--4302.

\bibitem[\protect\citeauthoryear{Li, Alderson, Doyle, and Willinger}{Li
  et~al.}{2005}]{li05:_towards_a_theory_of_scale_free_graphs}
Li, L., D.~Alderson, J.~C. Doyle, and W.~Willinger (2005).
\newblock Towards a theory of scale-free graphs: {D}efinition, properties, and
  implications.
\newblock {\em Internet Math.\/}~{\em 2}, 431--523.

\bibitem[\protect\citeauthoryear{Marchette and Priebe}{Marchette and
  Priebe}{2008}]{marchette2008pul}
Marchette, D.~J. and C.~E. Priebe (2008).
\newblock Predicting unobserved links in incompletely observed networks.
\newblock {\em Computat. Statist. Data Anal.\/}~{\em 52}, 1373--1386.

\bibitem[\protect\citeauthoryear{Massen and Doye}{Massen and
  Doye}{2006}]{massen06:_thermodynamics_of_community_structure}
Massen, C.~P. and J.~P.~K. Doye (2006).
\newblock Thermodynamics of community structure.
\newblock Unpublished manuscript, available online.

\bibitem[\protect\citeauthoryear{Moreno and Jennings}{Moreno and
  Jennings}{1938}]{moreno38:_statistics_of_social_configurations}
Moreno, J.~L. and H.~H. Jennings (1938).
\newblock Statistics of social configurations.
\newblock {\em Sociometry\/}~{\em 1}, 342--374.

\bibitem[\protect\citeauthoryear{Newman}{Newman}{2003}]{newman03:_structure_an%
d_function_of_comples_networks}
Newman, M. E.~J. (2003).
\newblock The structure and function of complex networks.
\newblock {\em SIAM Rev.\/}~{\em 45}, 167--256.

\bibitem[\protect\citeauthoryear{Newman}{Newman}{2006}]{newman06:_modularity_a%
nd_community_structure}
Newman, M. E.~J. (2006).
\newblock Modularity and community structure in networks.
\newblock {\em Proc. Natl. Acad. Sci. USA\/}~{\em 103}, 8577--8582.

\bibitem[\protect\citeauthoryear{Newman and Leicht}{Newman and
  Leicht}{2007}]{newman07:_mixture_models_and_exploratory_analysis}
Newman, M. E.~J. and E.~A. Leicht (2007).
\newblock Mixture models and exploratory analysis in networks.
\newblock {\em Proc. Natl. Acad. Sci. USA\/}~{\em 104}, 9564--9569.

\bibitem[\protect\citeauthoryear{Newman, Strogatz, and Watts}{Newman
  et~al.}{2001}]{newman01:_random_graphs_with_arbitrary_degree_distributions}
Newman, M. E.~J., S.~H. Strogatz, and D.~J. Watts (2001).
\newblock Random graphs with arbitrary degree distributions and their
  applications.
\newblock {\em Phys. Rev. E\/}~{\em 64}, 26118(1--17).

\bibitem[\protect\citeauthoryear{Palla, Der{\'e}nyi, Farkas, and Vicsek}{Palla
  et~al.}{2005}]{palla05:_uncovering_the_overlapping_community_structure_of_ne%
tworks}
Palla, G., I.~Der{\'e}nyi, I.~Farkas, and T.~Vicsek (2005).
\newblock Uncovering the overlapping community structure of complex networks in
  nature and society.
\newblock {\em Nature\/}~{\em 435}, 814--818.

\bibitem[\protect\citeauthoryear{Pemantle}{Pemantle}{2007}]{pemantle07:_survey%
_of_random_processes_with_reinforcement}
Pemantle, R. (2007).
\newblock A survey of random processes with reinforcement.
\newblock {\em Probab. Surv.\/}~{\em 4}, 1--79.

\bibitem[\protect\citeauthoryear{Pothen, Simon, and Liou}{Pothen
  et~al.}{1990}]{pothen90:_partitioning_sparse_matrices}
Pothen, A., H.~D. Simon, and K.-P. Liou (1990).
\newblock Partitioning sparse matrices with eigenvectors of graphs.
\newblock {\em SIAM J. Matrix Anal. Appl.\/}~{\em 11}, 430--452.

\bibitem[\protect\citeauthoryear{Radicchi, Castellano, Cecconi, Loreto, and
  Parisi}{Radicchi
  et~al.}{2004}]{radicchi04:_defining_and_identifying_communities_in_networks}
Radicchi, F., C.~Castellano, F.~Cecconi, V.~Loreto, and D.~Parisi (2004).
\newblock Defining and identifying communities in networks.
\newblock {\em Proc. Natl. Acad. Sci. USA\/}~{\em 101}, 2658--2663.

\bibitem[\protect\citeauthoryear{Reichardt and Bornholdt}{Reichardt and
  Bornholdt}{2004}]{reichardt04:_detecting_fuzzy_community_structure}
Reichardt, J. and S.~Bornholdt (2004).
\newblock Detecting fuzzy community structures in complex networks with a
  {P}otts model.
\newblock {\em Phys. Rev. Lett.\/}~{\em 93}, 218701(1--4).

\bibitem[\protect\citeauthoryear{Rosvall and Bergstrom}{Rosvall and
  Bergstrom}{2007}]{rosvall07:_an_information_theoretic_framework_for_resolvin%
g_community_structure}
Rosvall, M. and C.~T. Bergstrom (2007).
\newblock An information-theoretic framework for resolving community structure
  in complex networks.
\newblock {\em Proc. Natl. Acad. Sci. USA\/}~{\em 104}, 7327--7331.

\bibitem[\protect\citeauthoryear{Rosvall and Bergstrom}{Rosvall and
  Bergstrom}{2008}]{rosvall08:_maps_of_random_walks_on_complex_networks}
Rosvall, M. and C.~T. Bergstrom (2008).
\newblock Maps of random walks on complex networks reveal community structure.
\newblock {\em Proc. Natl. Acad. Sci. USA\/}~{\em 105}, 1118--1123.

\bibitem[\protect\citeauthoryear{Snijders}{Snijders}{1981}]{snijders81:_the_de%
gree_variance}
Snijders, T. A.~B. (1981).
\newblock The degree variance: {A}n index of graph heterogeneity.
\newblock {\em Social Networks\/}~{\em 3}, 163--223.

\bibitem[\protect\citeauthoryear{Snijders, Pattison, Robins, and
  Handcock}{Snijders et~al.}{2006}]{snijders06:_new_specifications_for_ERGMs}
Snijders, T. A.~B., P.~E. Pattison, G.~L. Robins, and M.~S. Handcock (2006).
\newblock New specifications for exponential random graph models.
\newblock {\em Sociolog. Methodol.\/}~{\em 36}, 99--153.

\bibitem[\protect\citeauthoryear{Stutzbach, Rejaie, Duffield, Sen, and
  Willinger}{Stutzbach
  et~al.}{2006}]{willinger06:_unbiased_sampling_peer_to_peer}
Stutzbach, D., R.~Rejaie, N.~Duffield, S.~Sen, and W.~Willinger (2006).
\newblock On unbiased sampling for unstructured peer-to-peer networks.
\newblock In {\em Proc. 6th ACM SIGCOMM Conference on Internet Measurement},
  pp.\  27--40.

\bibitem[\protect\citeauthoryear{Thompson}{Thompson}{2006}]{thompson06:_adapti%
ve_web_sampling}
Thompson, S.~K. (2006).
\newblock Adaptive web sampling.
\newblock {\em Biometrics\/}~{\em 62}, 1224--1234.

\bibitem[\protect\citeauthoryear{Viger and Latapy}{Viger and
  Latapy}{2005}]{viger2005eas}
Viger, F. and M.~Latapy (2005).
\newblock Efficient and simple generation of random simple connected graphs
  with prescribed degree sequence.
\newblock In L.~Wang (Ed.), {\em Proc. 11th Annual International Computing and
  Combinatorics Conference}, pp.\  440--449. Berlin: Springer.

\bibitem[\protect\citeauthoryear{von Luxburg}{von
  Luxburg}{2007}]{luxburg2007tsc}
von Luxburg, U. (2007).
\newblock A tutorial on spectral clustering.
\newblock {\em Statist. Comput.\/}~{\em 17}, 395--416.

\bibitem[\protect\citeauthoryear{von Luxburg, Belkin, and Bousquet}{von Luxburg
  et~al.}{2008}]{vonluxburg2008csc}
von Luxburg, U., M.~Belkin, and O.~Bousquet (2008).
\newblock Consistency of spectral clustering.
\newblock {\em Ann. Statist.\/}~{\em 36}, 555--586.

\bibitem[\protect\citeauthoryear{Wang, Wang, Zhang, and Chen}{Wang
  et~al.}{2007}]{wang07:_detecting_community_structure_by_optimal_rearrangemen%
t_clustering}
Wang, R., Y.~Wang, X.~Zhang, and L.~Chen (2007).
\newblock Detecting community structure in complex networks by optimal
  rearrangement clustering.
\newblock In J.~G. Carbonell and J.~Siekmann (Eds.), {\em Papers from the PAKDD
  2007 International Workshops}, pp.\  119--130. Berlin: Springer.

\bibitem[\protect\citeauthoryear{Wang and Wong}{Wang and
  Wong}{1987}]{wang87:_stochastic_blockmodels_for_directed_graphs}
Wang, Y.~J. and G.~Y. Wong (1987).
\newblock Stochastic blockmodels for directed graphs.
\newblock {\em J. Am. Statist. Ass.\/}~{\em 82}, 8--19.

\bibitem[\protect\citeauthoryear{Wasserman and Faust}{Wasserman and
  Faust}{1994}]{wasserman94:_social_network_analysis}
Wasserman, S. and K.~Faust (1994).
\newblock {\em Social Network Analysis: {M}ethods and Applications}.
\newblock Cambridge University Press.

\bibitem[\protect\citeauthoryear{Watts and Strogatz}{Watts and
  Strogatz}{1998}]{watts98:_collective_dynamics_of_small_world_networks}
Watts, D.~J. and S.~H. Strogatz (1998).
\newblock Collective dynamics of `small-world' networks.
\newblock {\em Nature\/}~{\em 393}, 440--442.

\bibitem[\protect\citeauthoryear{White and Smyth}{White and
  Smyth}{2005}]{white05:_spectral_clustering_approach_to_finding_communities}
White, S. and P.~Smyth (2005).
\newblock A spectral clustering approach to finding communities in graphs.
\newblock In {\em Proc. SIAM International Conference on Data Mining}, pp.\
  274--285.

\bibitem[\protect\citeauthoryear{Zachary}{Zachary}{1977}]{zachary77:_an_inform%
ation_flow_model_for_conflict_in_small_groups}
Zachary, W.~W. (1977).
\newblock An information flow model for conflict and fission in small groups.
\newblock {\em J. Anthropolog. Res.\/}~{\em 33}, 452--473.

\bibitem[\protect\citeauthoryear{Zheng, Salganik, and Gelman}{Zheng
  et~al.}{2006}]{zheng06:_how_many_people_do_you_know_in_prison}
Zheng, T., M.~J. Salganik, and A.~Gelman (2006).
\newblock How many people do you know in prison?: {U}sing overdispersion in
  count data to estimate social structure in networks.
\newblock {\em J. Am. Statist. Ass.\/}~{\em 101}, 409--423.

\end{thebibliography}

\end{document}